\documentclass[a4paper,11pt]{article}

\usepackage{contribution}

\newcommand{\weblink}[2][]{%
    \ifthenelse{\equal{#1}{}}%
    {\textnormal{\url{#2}}}%
    {\textnormal{\href{#2}{#1}}}%
}

\newcommand{\acknowledgements}[1]{%
  \bigskip\bigskip
  \textsf{\textbf{\Large Acknowledgements}} \\[2ex]
  {#1}
  \bigskip
}

\def\beq{\begin{equation}}
\def\eeq#1{\label{#1}\end{equation}}
\def\eeqn{\end{equation}}

\def\beqa{\begin{eqnarray}}
\def\eeqa#1{\label{#1}\end{eqnarray}}
\def\eeqan{\end{eqnarray}}

\let\bar=\overbar

\def\Dslash{\not{\hbox{\kern-4pt $D$}}}
\def\dslash{\not{\hbox{\kern-2pt $\del$}}}

\def\msb{{\bar{\ssstyle M \kern -1pt S}}}

\newcommand{\contribution}[7][]{%
  \clearpage
  \thispagestyle{plain}
  \ifthenelse{\equal{#1}{}}
  {\hypersetup{pdftitle={#2}}}
  {\hypersetup{pdftitle={#1}}}
  \hypersetup{pdfauthor={{#3} {#4}}}
  {\centering\normalfont\LARGE\bfseries\sffamily #2 \par\nobreak}
  \lhead{}
  \chead{%
    \textit{\footnotesize XIV International Conference on Hadron Spectroscopy
      (\weblink[\textit{hadron2011}]{http://www.hadron2011.de}), 13-17 June 2011, Munich, Germany}%
  }
  \rhead{}
  \bigskip
  \begin{center}
    {#3} {#4}\ifthenelse{\equal{#6}{}}{}{\footnote{\weblink[#6]{mailto:#6}}}
    \ifthenelse{\equal{#7}{}}{}{#7} \\
    \textit{#5}
  \end{center}
  \bigskip
}

\renewcommand{\abstract}[1]{%
  \begin{center}
    \begin{minipage}{0.85\textwidth}
      \begin{footnotesize}
        #1
      \end{footnotesize}
    \end{minipage}
  \end{center}
  \bigskip
}

\begin{document}

%
%
%
%
%
{  


%

\contribution[Recent developments in quarkonium and open flavour production calculations]  
{Recent developments in quarkonium and open flavour \\ production calculations}  
{Mathias}{Butenschoen}  
{{II.} Institut f\"ur Theoretische Physik, Universit\"at Hamburg \\
Luruper Chaussee 149, 22761 Hamburg, Germany}  
{mathias.butenschoen@desy.de}  
{}  
%

\abstract{%
  This report reviews recent theory progress in the field of heavy quarkonium and open heavy flavour production calculations.
}
%

\section{\label{sec:heavyquarkonium}Heavy Quarkonium Production}

\subsection{Introduction}

Heavy quarkonia are bound states of a heavy quark and its antiquark. There are charmonia and bottomonia. According to the factorization theorem of nonrelativistic QCD (NRQCD) \cite{Bodwin:1994jh}, the cross section to produce a heavy quarkonium $H$ factorizes according to
\begin{equation}
 \sigma(ab\to H+X) = \sum_n \sigma(ab\to c\overline{c}[n] + X) \langle{\cal O}^{H}[n]\rangle,
\end{equation}
where the $\sigma(ab\to c\overline{c}[n] + X)$ are perturbatively calculated short distance cross sections describing the production of a heavy quark pair (here $c\overline{c}$) in an intermediate Fock state $n$, which does not have to be color neutral. The $\langle{\cal O}^{H}[n]\rangle$ are nonperturbative long distance matrix elements (LDMEs) extracted from experiment and describing the transition of that intermediate $c\overline{c}$ state into the physical $H$ via soft gluon radiation. NRQCD predicts each of the LDMEs to scale with a definite power of the relative heavy quark velocity $v\ll 1$, which serves as an additional expansion parameter besides $\alpha_s$: In case of $H=J/\psi$, the leading order contribution in the $v$ expansion stems from $n={^3S}_1^{[1]}$ and equals the traditional color singlet model (CSM) prediction, while the leading relativistic corrections are made up by the $^1S_0^{[8]}$, $^3S_1^{[8]}$, and $^3P_J^{[8]}$ states. The upper index ``8'' stands for color octet (CO), and these contributions are usually just called {\em the} color octet contributions. The CSM alone is theoretically incomplete due to uncancelled infrared divergences in the case of $p$ wave quarkonia. On the other hand, however, the validity of the NRQCD factorization and the universality of the LDMEs are still not proven and subject to dispute. Most of the work reviewed in the following therefore just aims at testing them.

\subsection{\label{sec:COCals}NLO calculations of color octet contributions}

The calculation of next-to-leading order (NLO) corrections to the short distance cross sections of the intermediate CO states, especially to the 
$^3P_J^{[8]}$ states, have proven challenging. But as for unpolarized $J/\psi$ production, up to now they have been calculated for all relevant collision processes. The $2\to 1$ processes for photo- and hadroproduction have already been calculated in 1998 \cite{Petrelli:1997ge}. Inclusive production in direct two photon collisions followed in 2005 \cite{Klasen:2004az}, in direct photoproduction \cite{Butenschoen:2009zy} and
electron-positron scattering neglecting the small $^3S_1^{[8]}$ contribution \cite{Zhang:2009ym} in 2009. The hadroproduction calculations \cite{Gong:2008ft} were still missing the $^3P_J^{[8]}$ contributions. Full calculations involving all CO states followed in 2010 with two independent works \cite{Ma:2010yw,Butenschoen:2010rq}. The missing pieces of single and double resolved two photon scattering, resolved photoproduction and the $^3S_1^{[8]}$ contributions of electron-positron scattering were finally presented in 2011 \cite{Butenschoen:2011yh}.

The two hadroproduction works \cite{Ma:2010yw,Butenschoen:2010rq} initially stirred some confusion, because the extracted CO LDMEs seem incompatible although the short distance cross sections agree within the expected numerical uncertainties. That difference is mainly due to the fact that in \cite{Butenschoen:2010rq} a combined fit to the transverse momentum $p_T$ distributions in H1 HERA photoproduction and CDF Tevatron hadroproduction data was performed, while in \cite{Ma:2010yw} a Tevatron-only fit was performed. When fitting to hadroproduction data alone, the fit is unconstrained, so only two linear combinations of the CO LDMEs can be extracted in \cite{Ma:2010yw}, and the fit results depend strongly on parameters like the lower cut on $p_T$. But when both groups perform the fit in the same way, meaning doing a three-parameter fit neglecting feed-down contributions like in \cite{Butenschoen:2010rq}, but fitting only the seven data points from the CDF Tevatron run 2 measurement \cite{Acosta:2004yw} with $p_T>7$~GeV like in \cite{Ma:2010yw}, the fit results do agree within the fit errors. So there is no obvious inconsistency between the two works.

\subsection{Global fit of J/$\psi$ CO LDMEs to unpolarized production data}

\begin{figure}
\includegraphics[width=3.75cm]{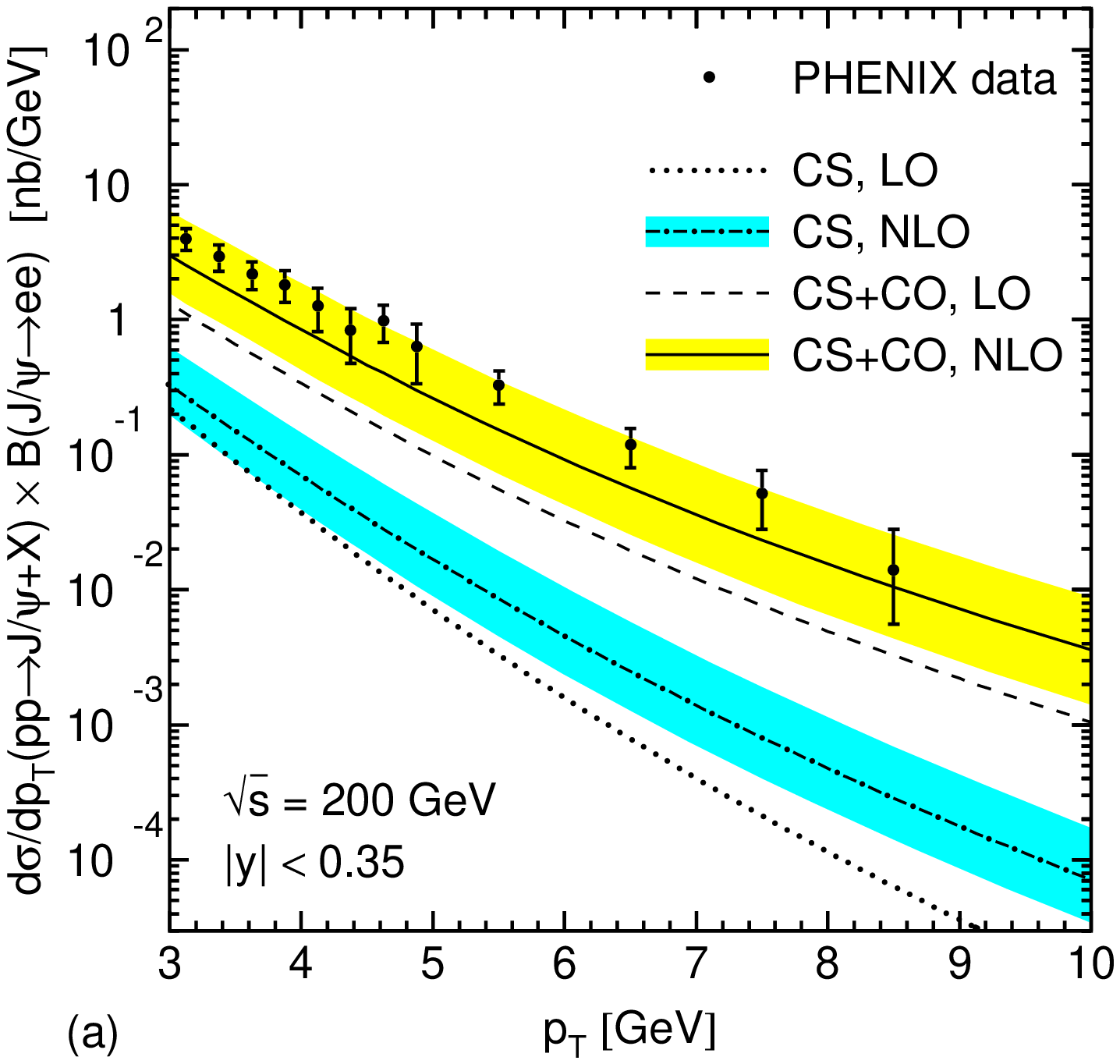}
\includegraphics[width=3.75cm]{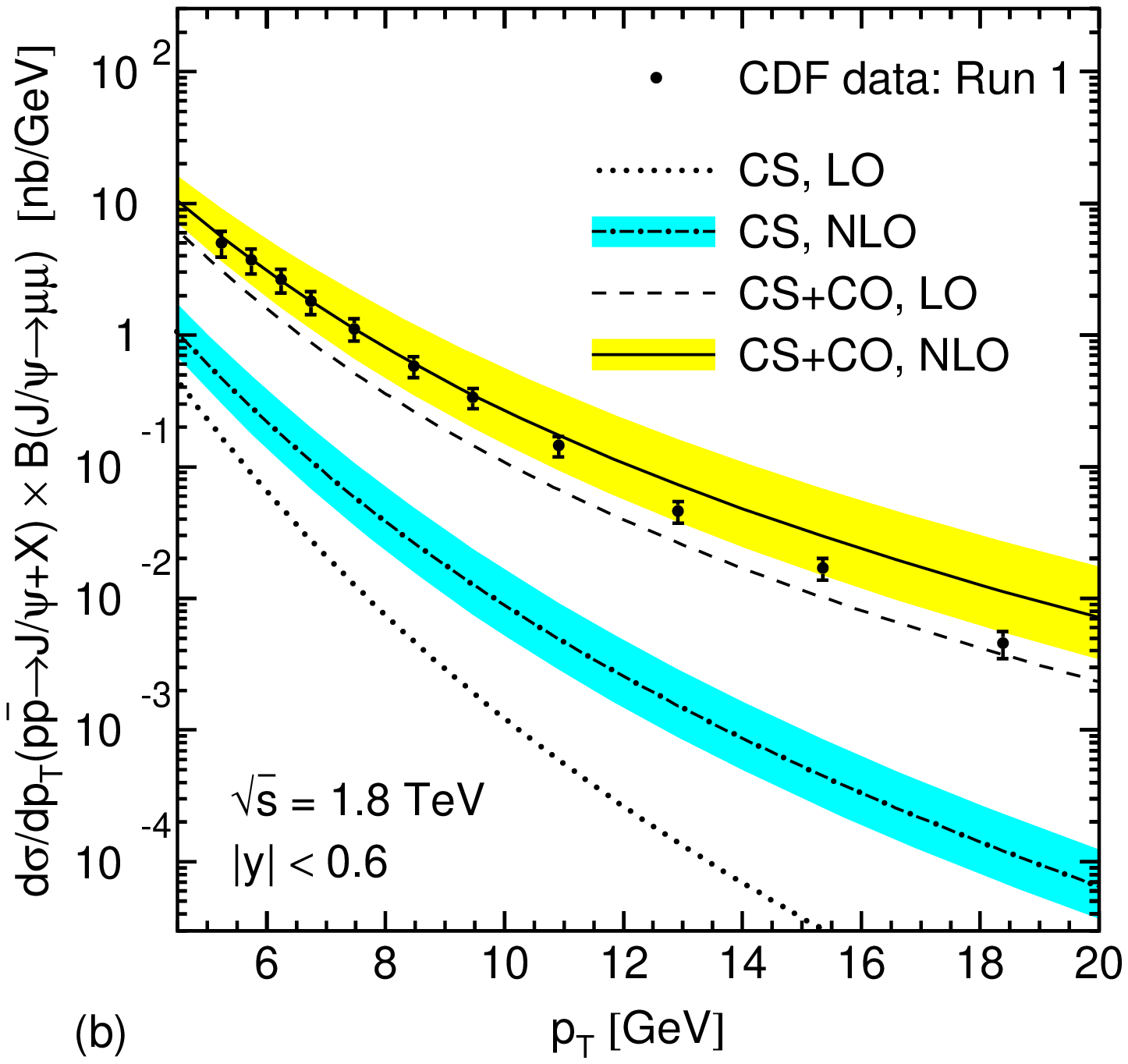}
\includegraphics[width=3.75cm]{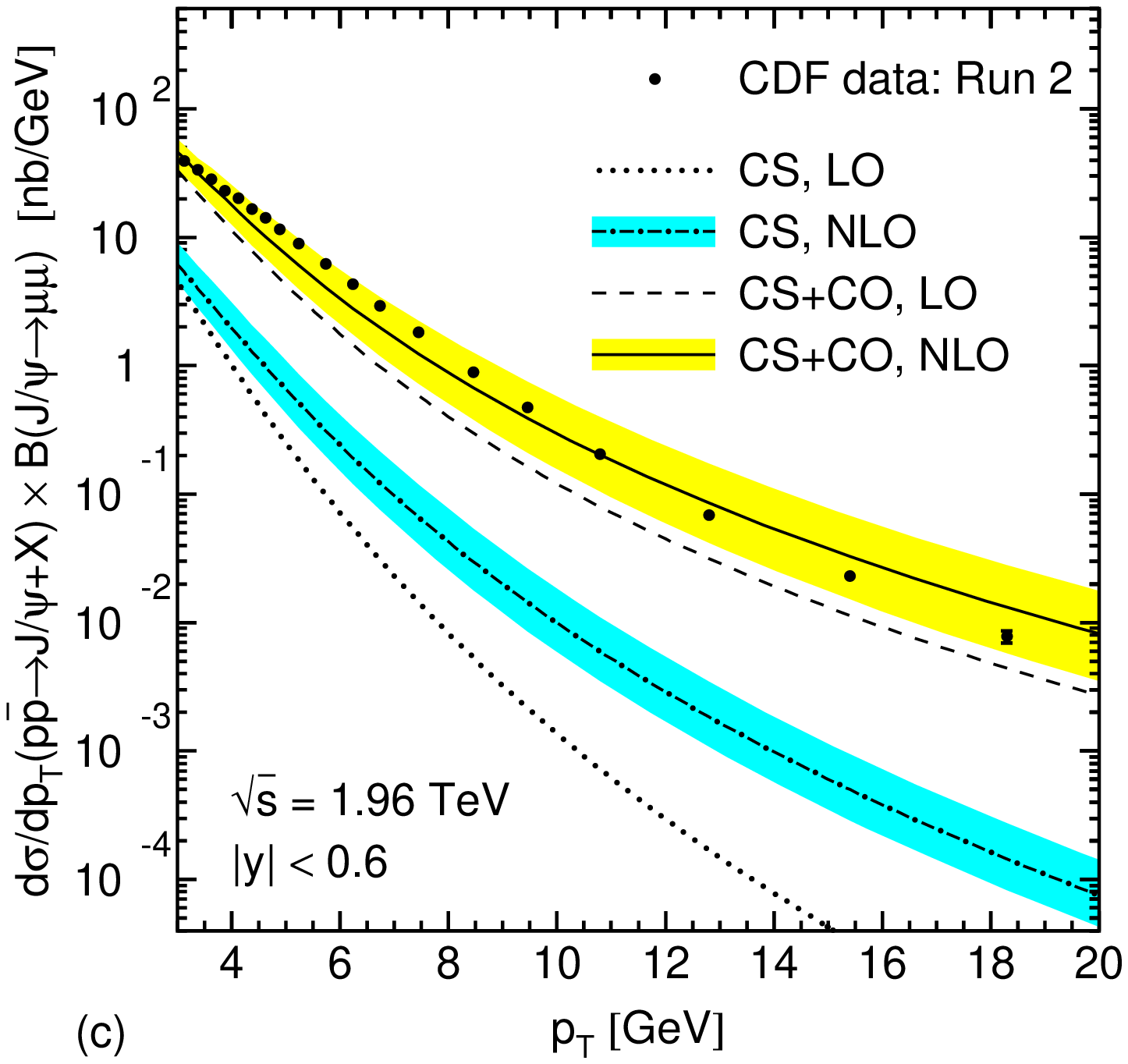}
\includegraphics[width=3.75cm]{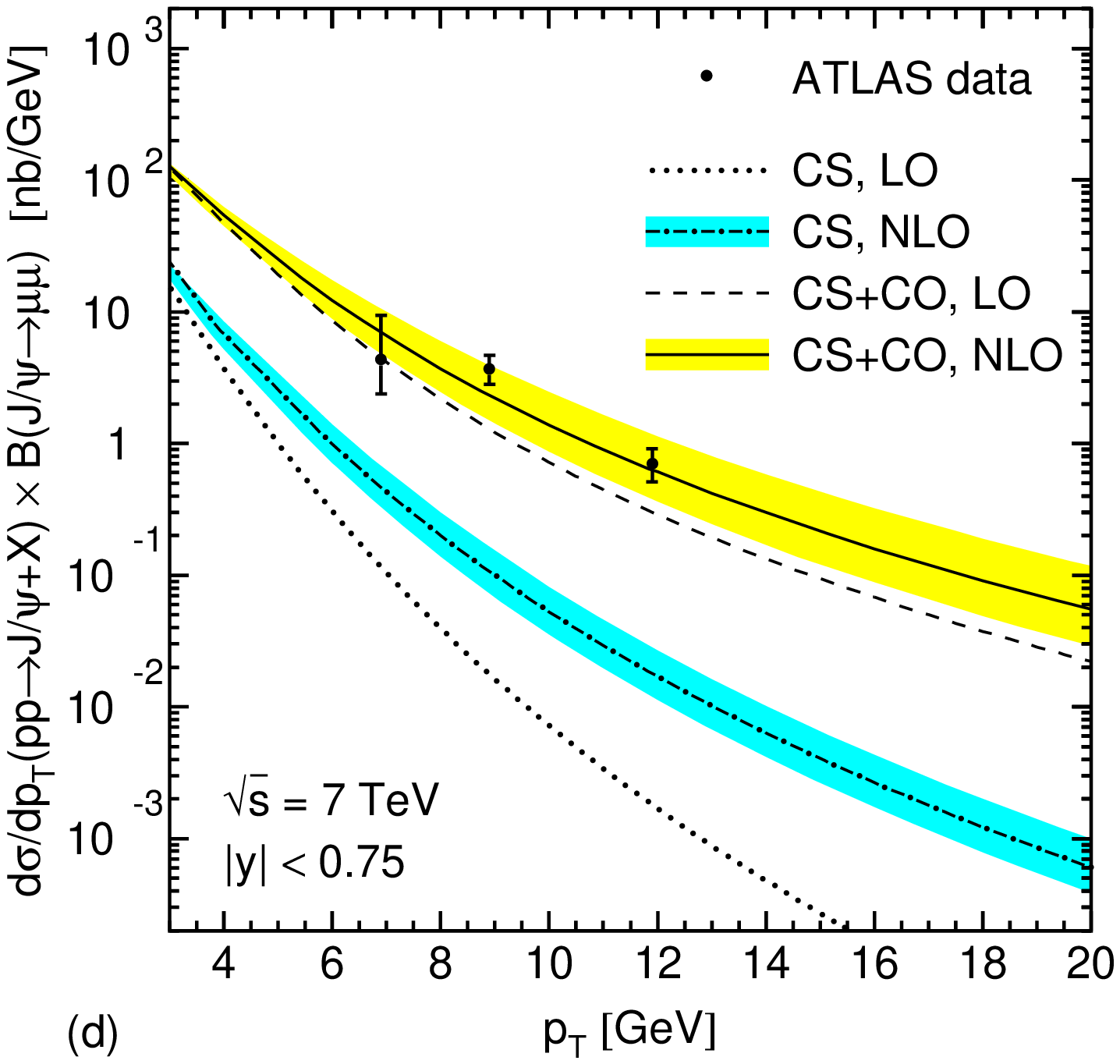}

\vspace{2pt}
\includegraphics[width=3.75cm]{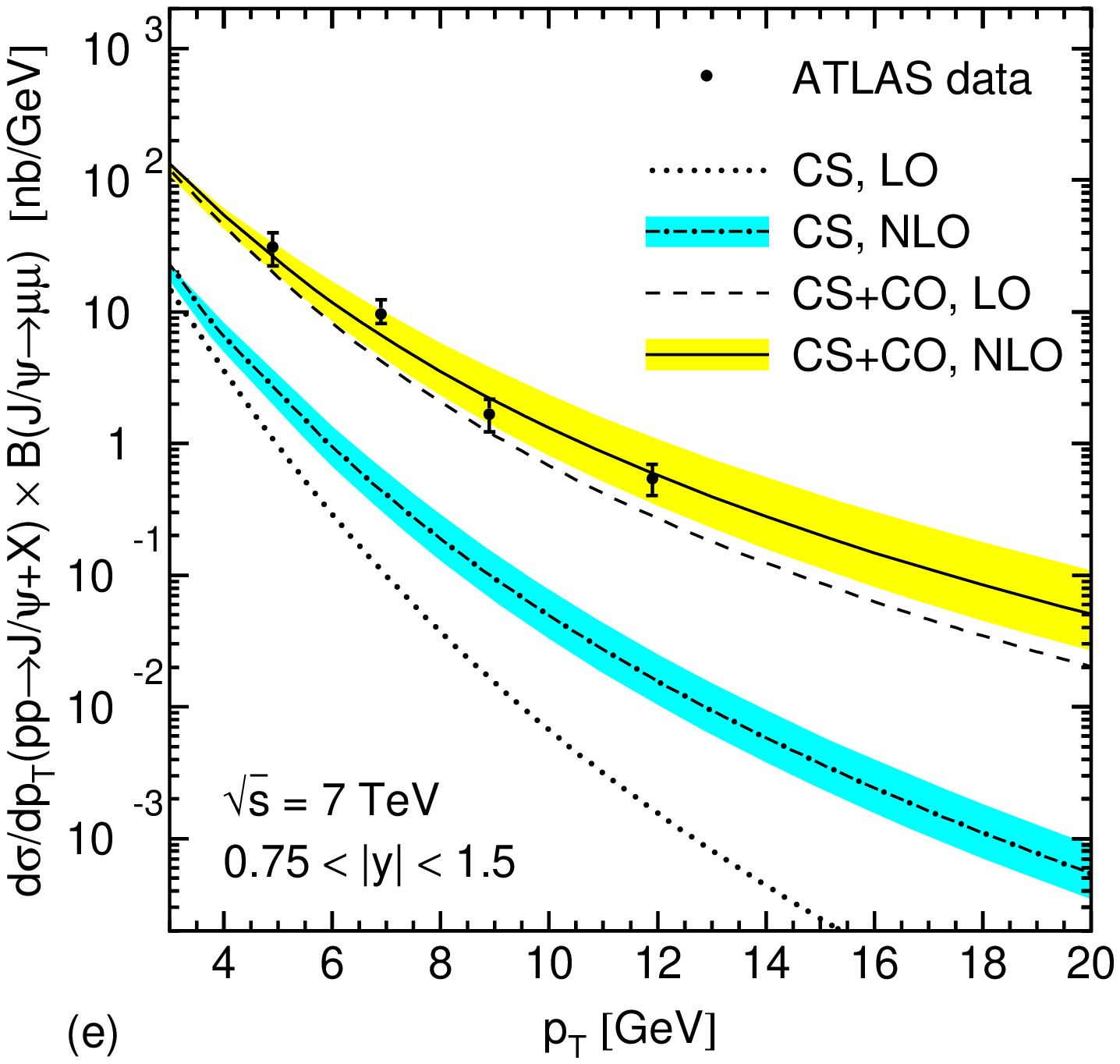}
\includegraphics[width=3.75cm]{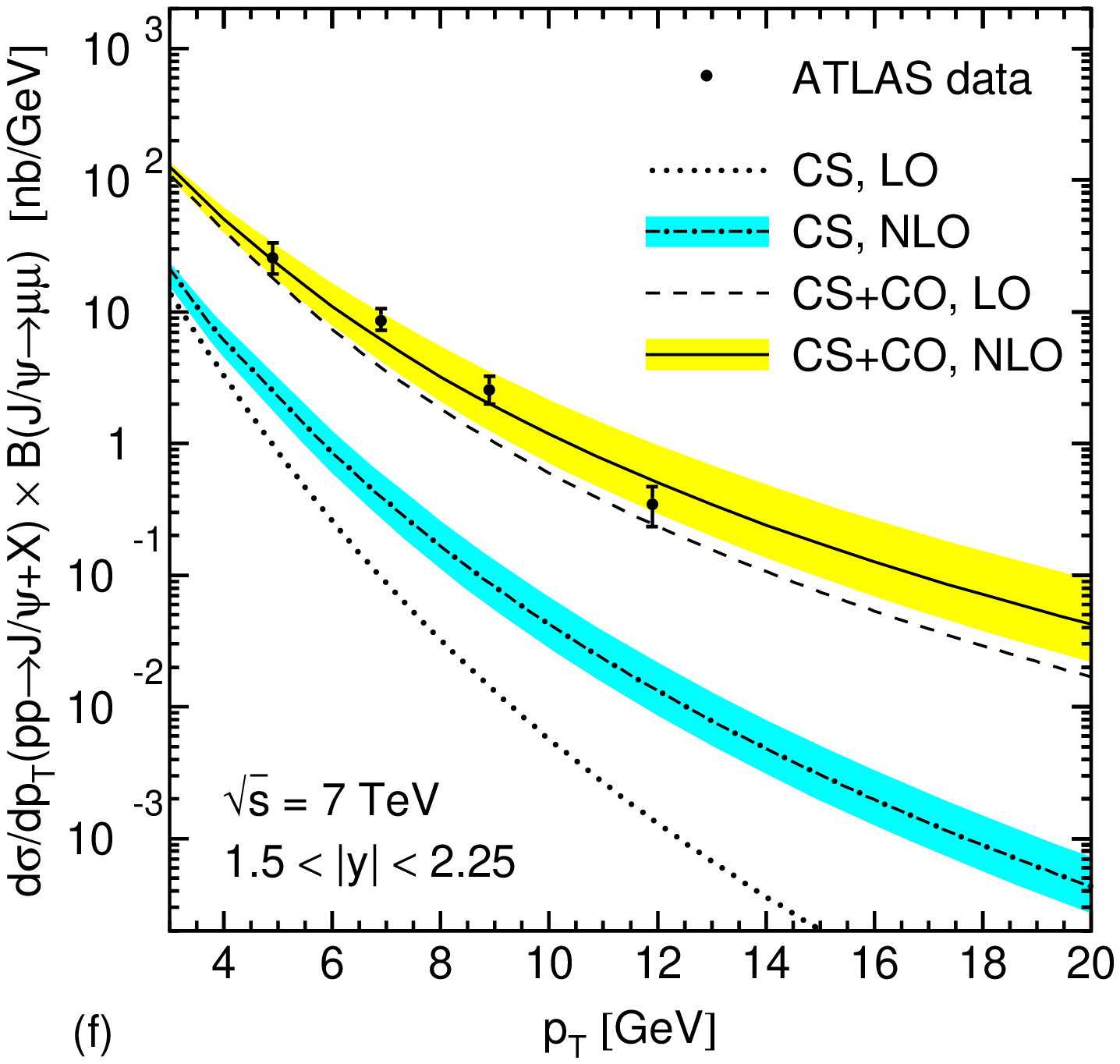}
\includegraphics[width=3.75cm]{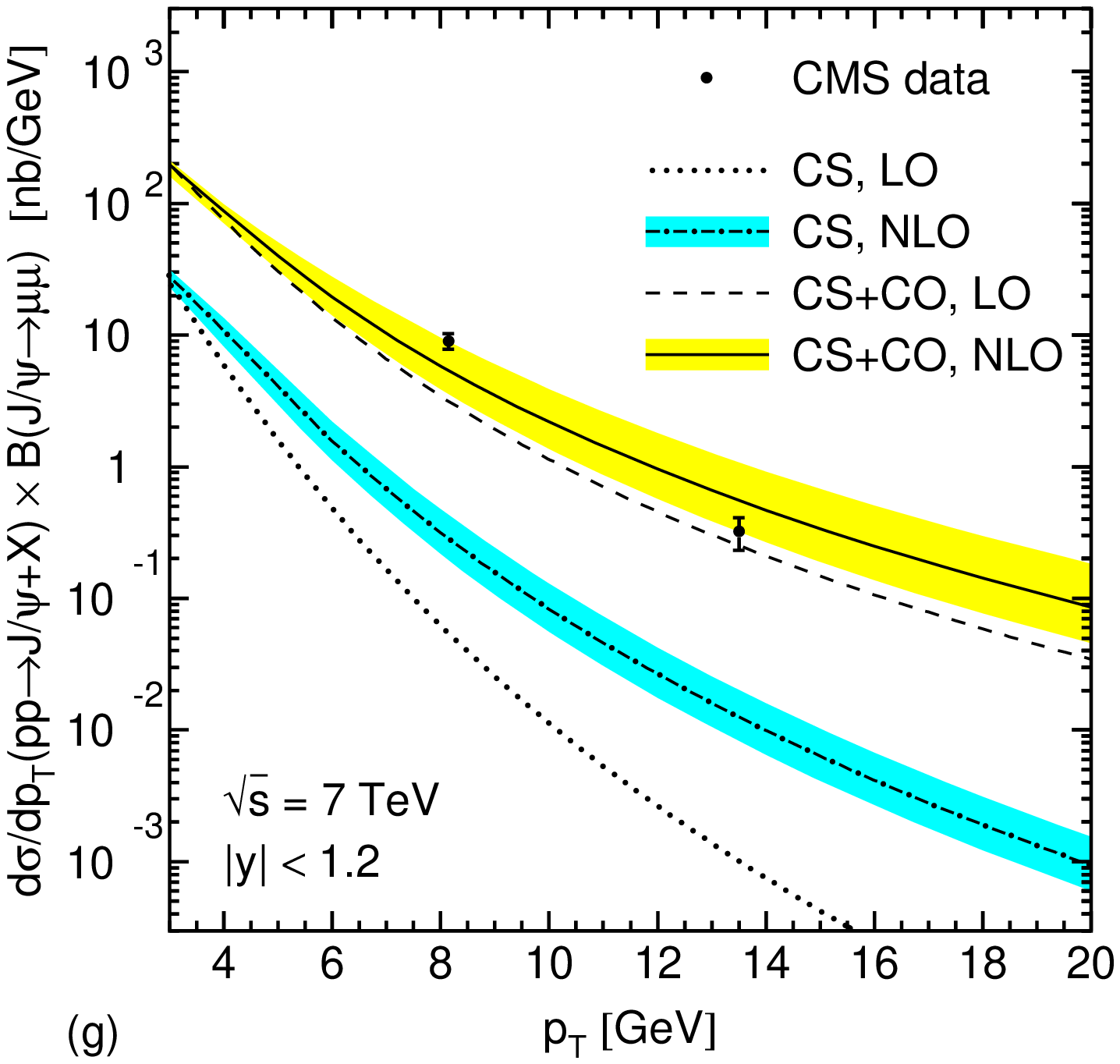}
\includegraphics[width=3.75cm]{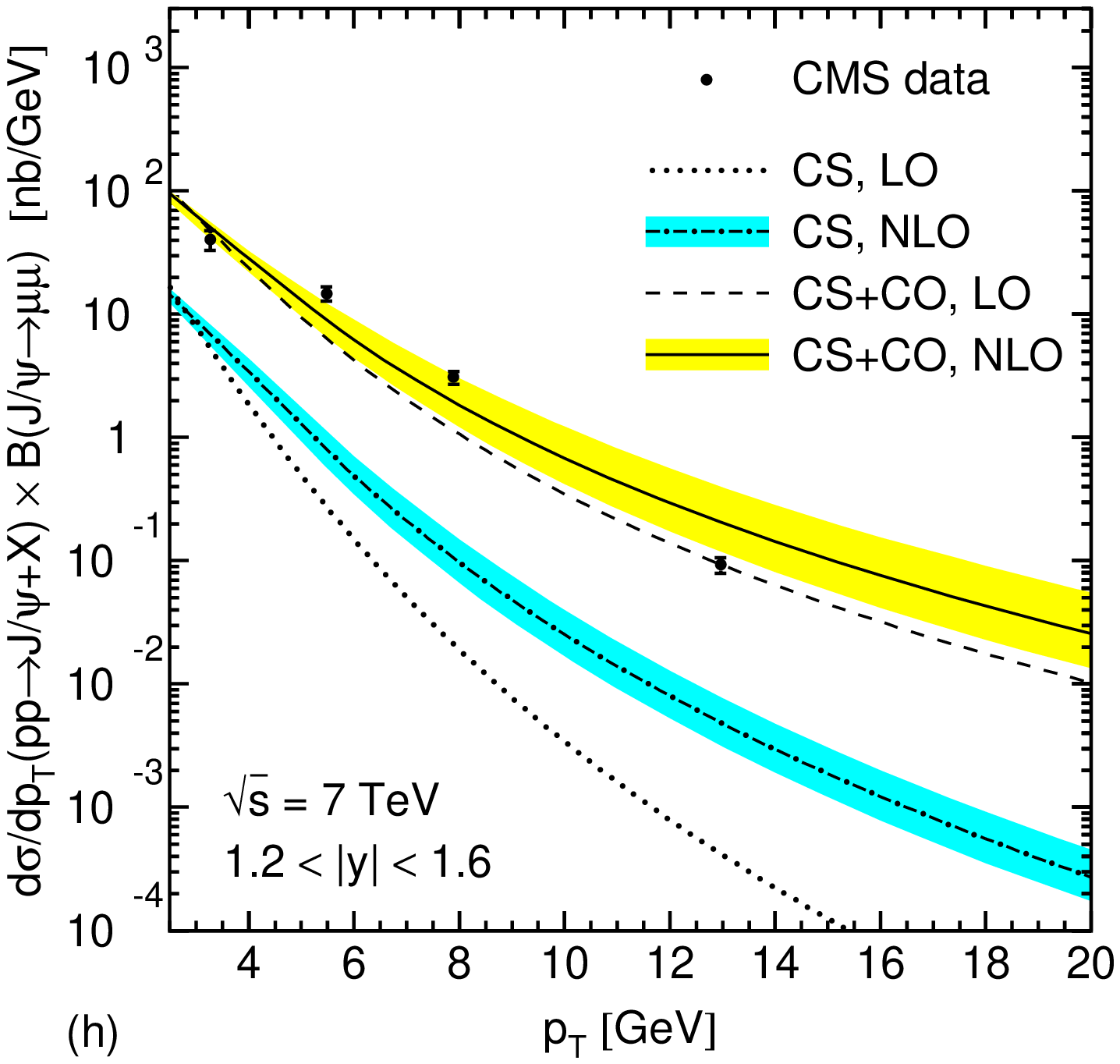}

\vspace{2pt}
\includegraphics[width=3.75cm]{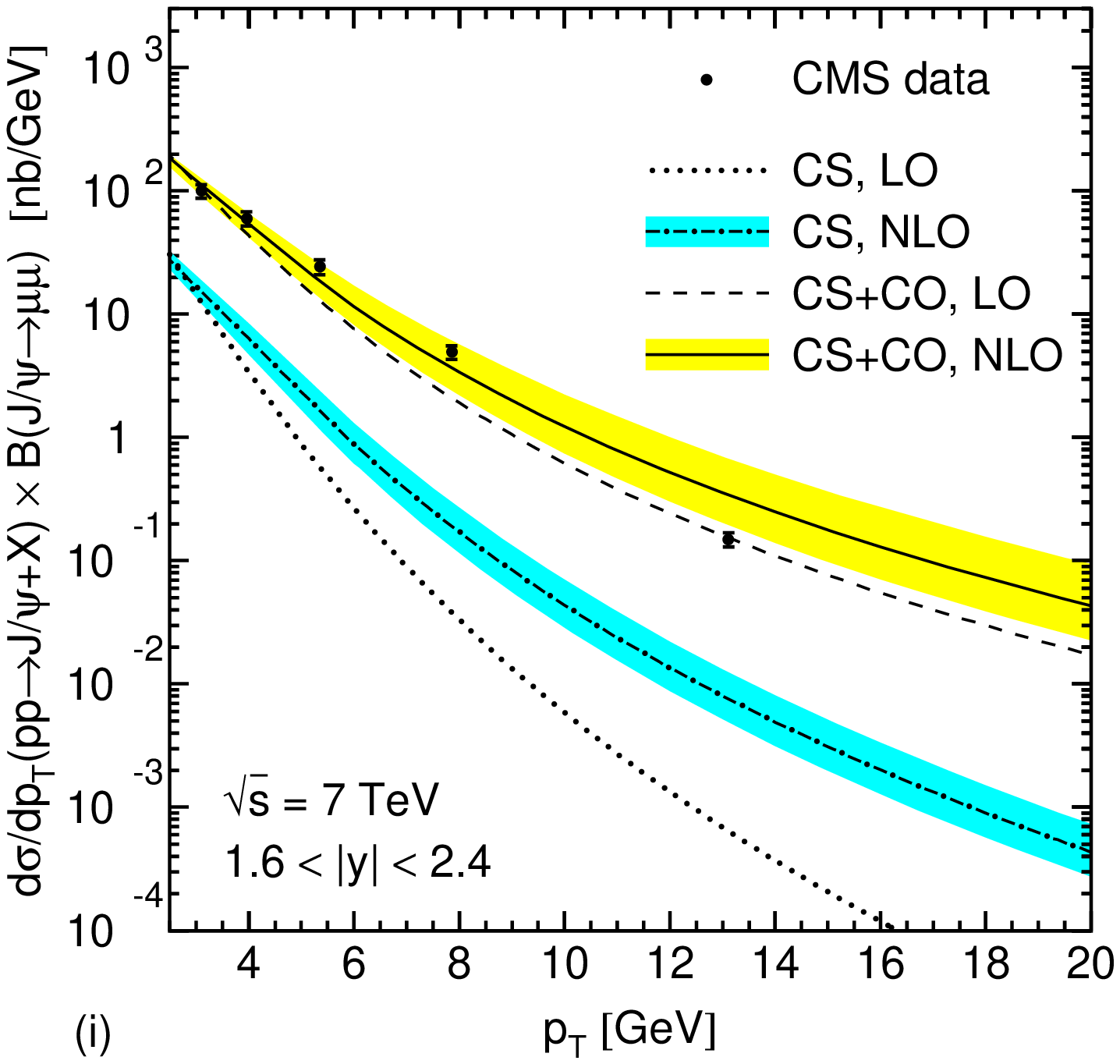}
\includegraphics[width=3.75cm]{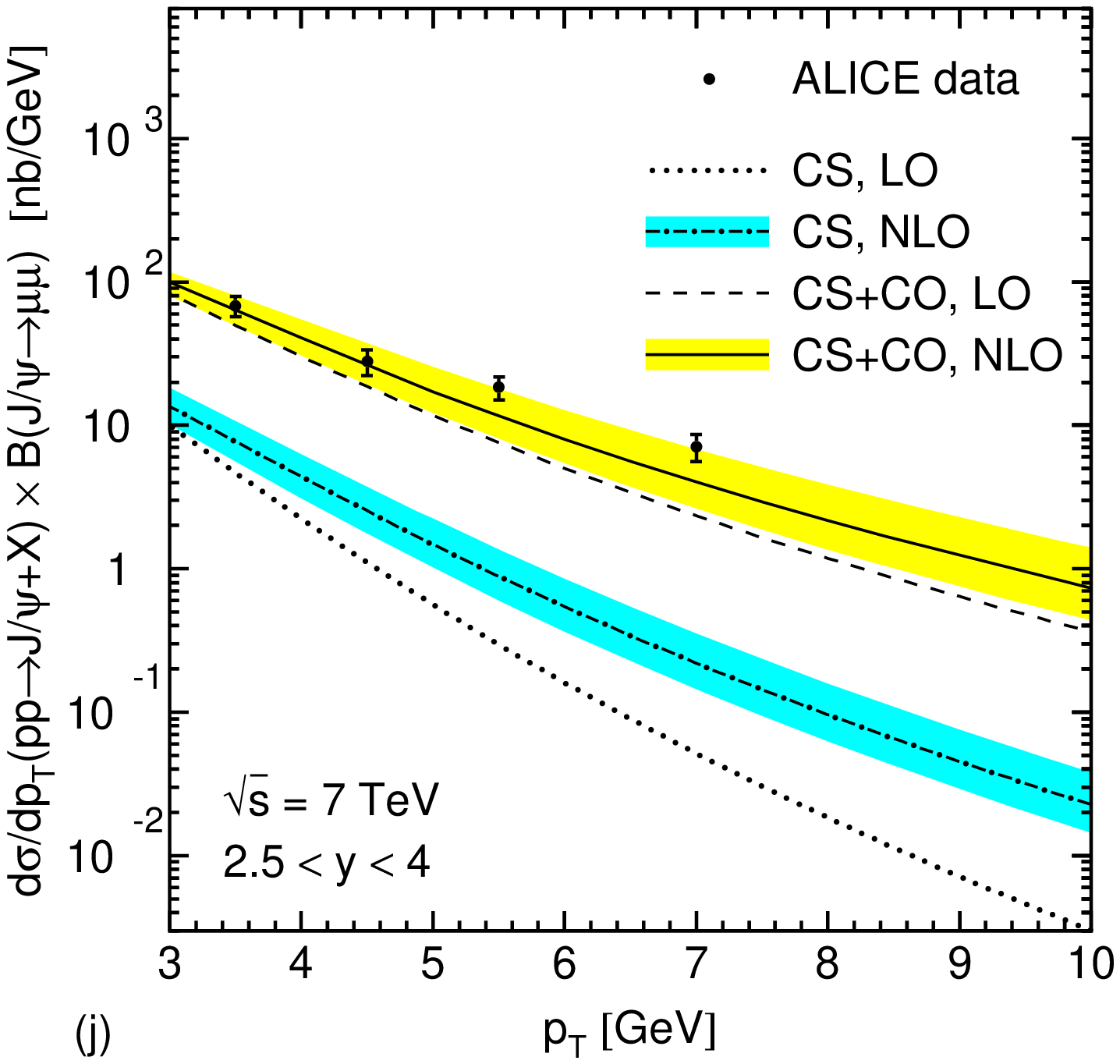}
\includegraphics[width=3.75cm]{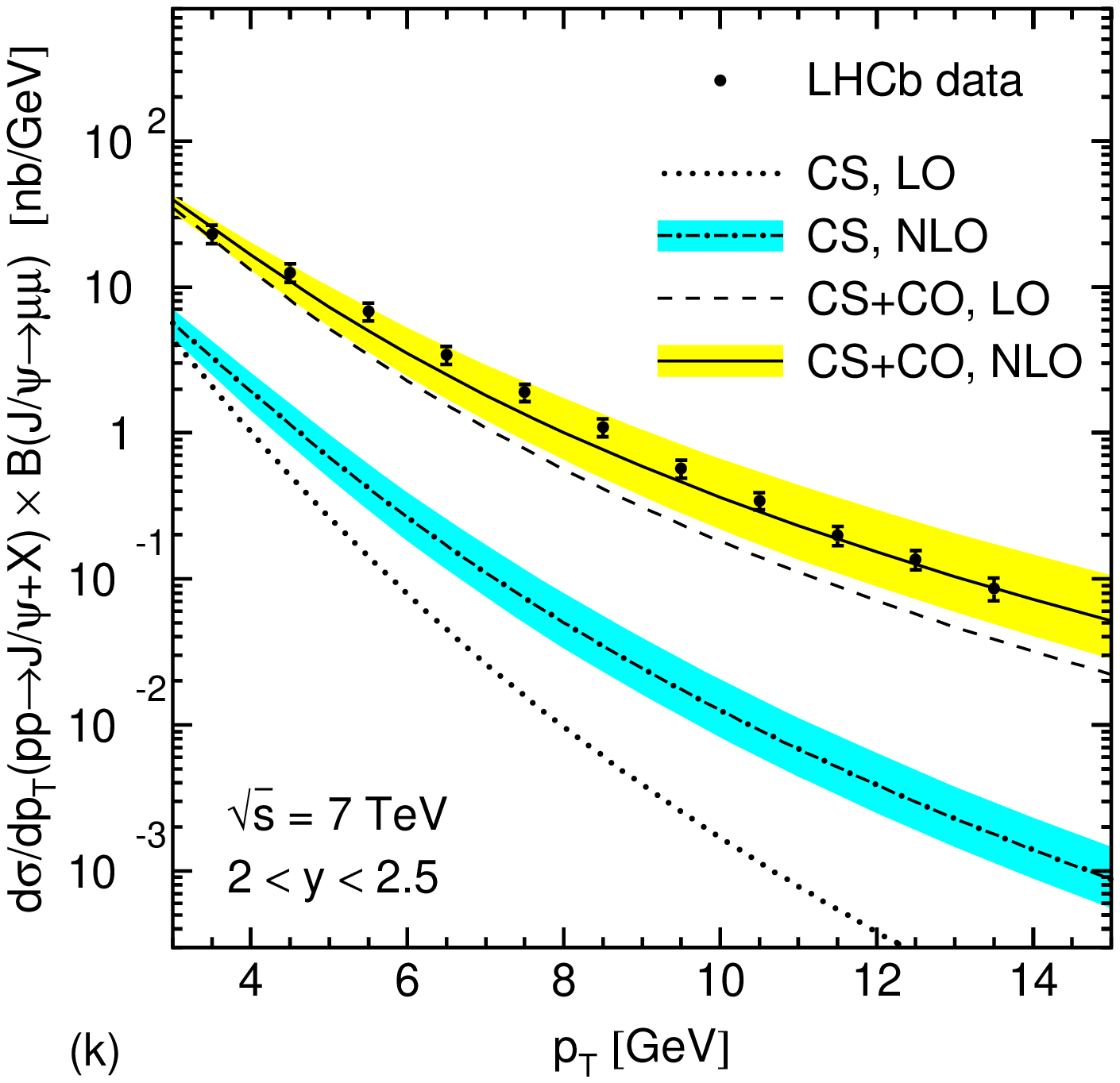}
\includegraphics[width=3.75cm]{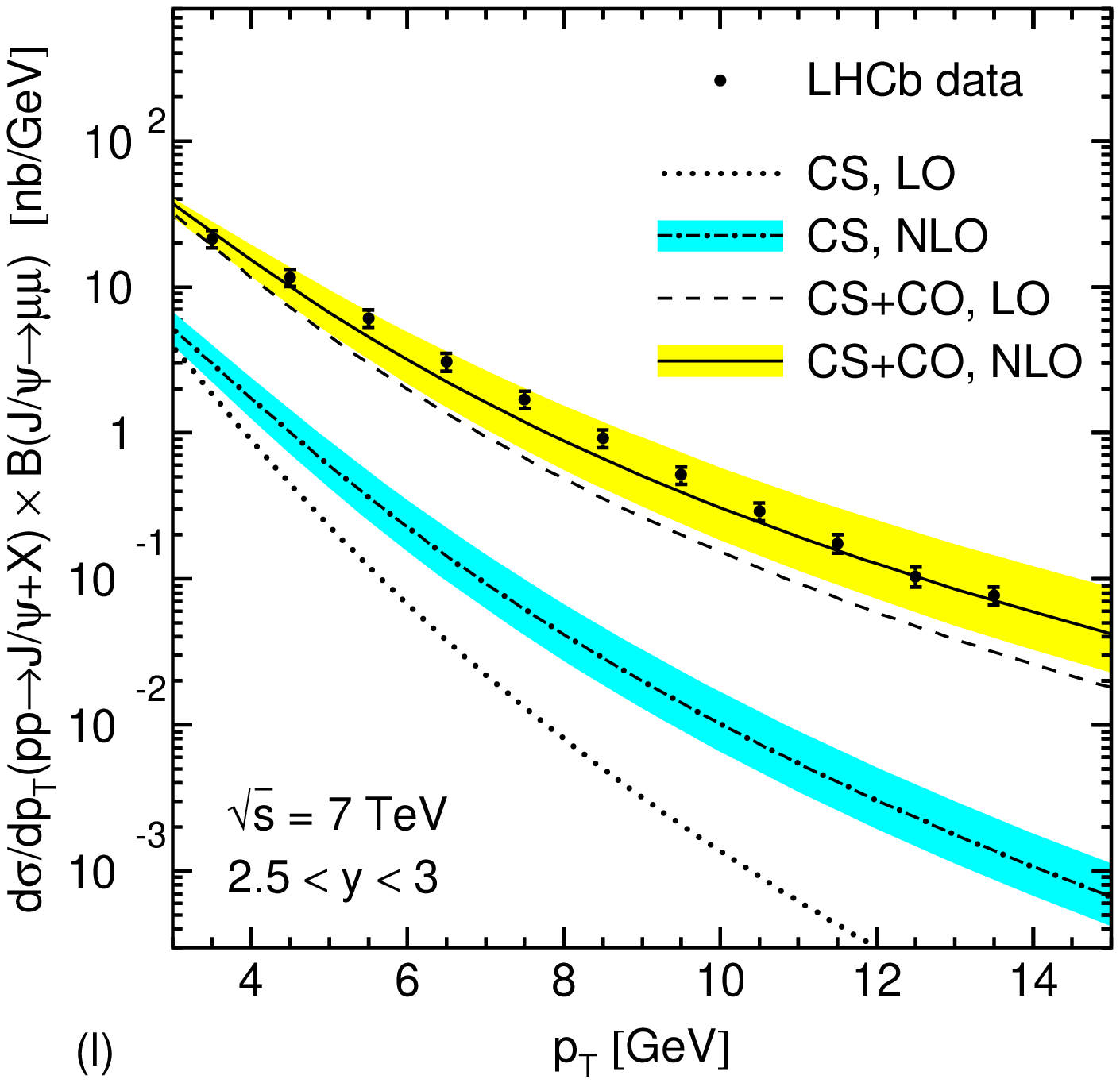}

\vspace{2pt}
\includegraphics[width=3.75cm]{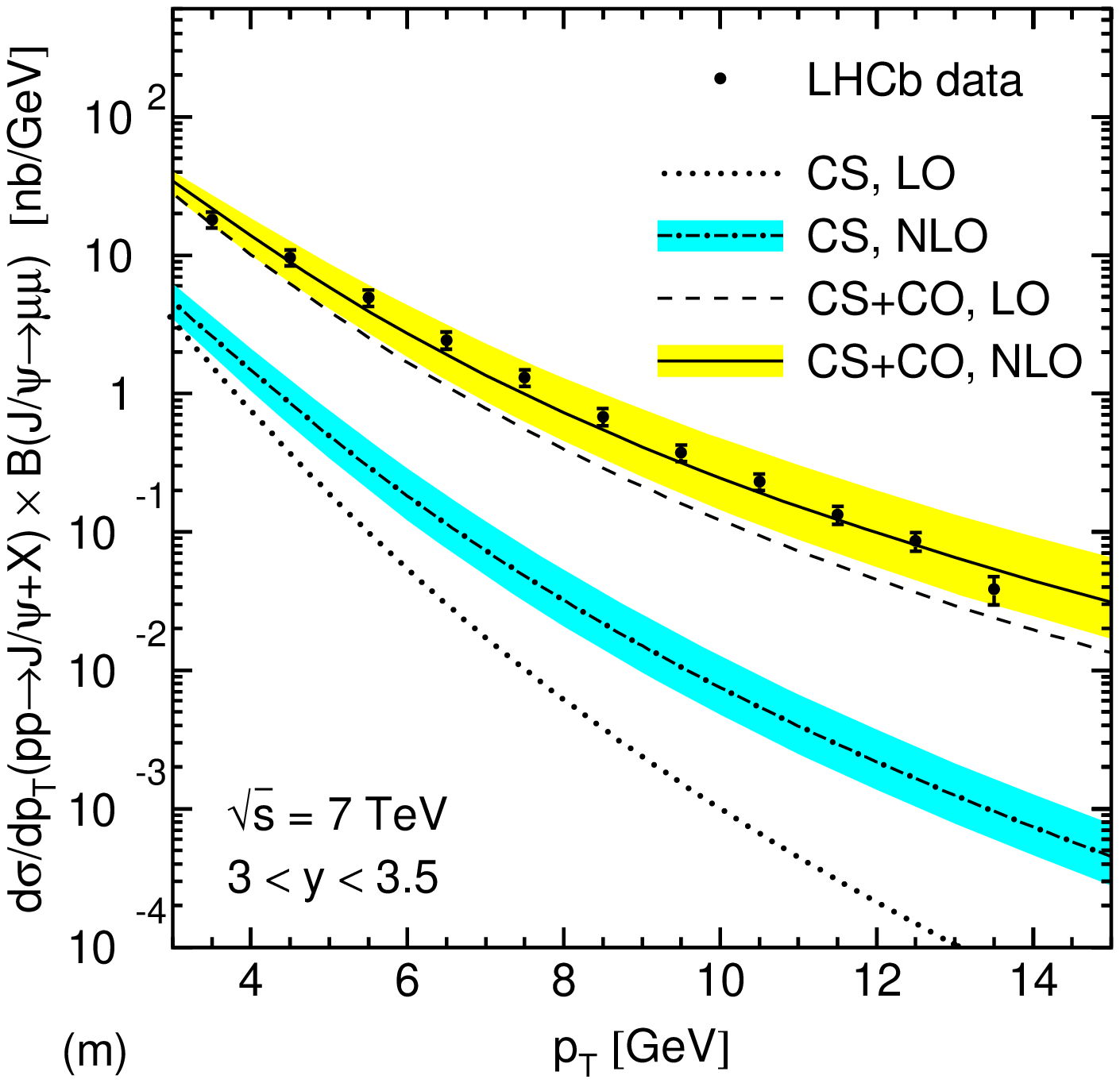}
\includegraphics[width=3.75cm]{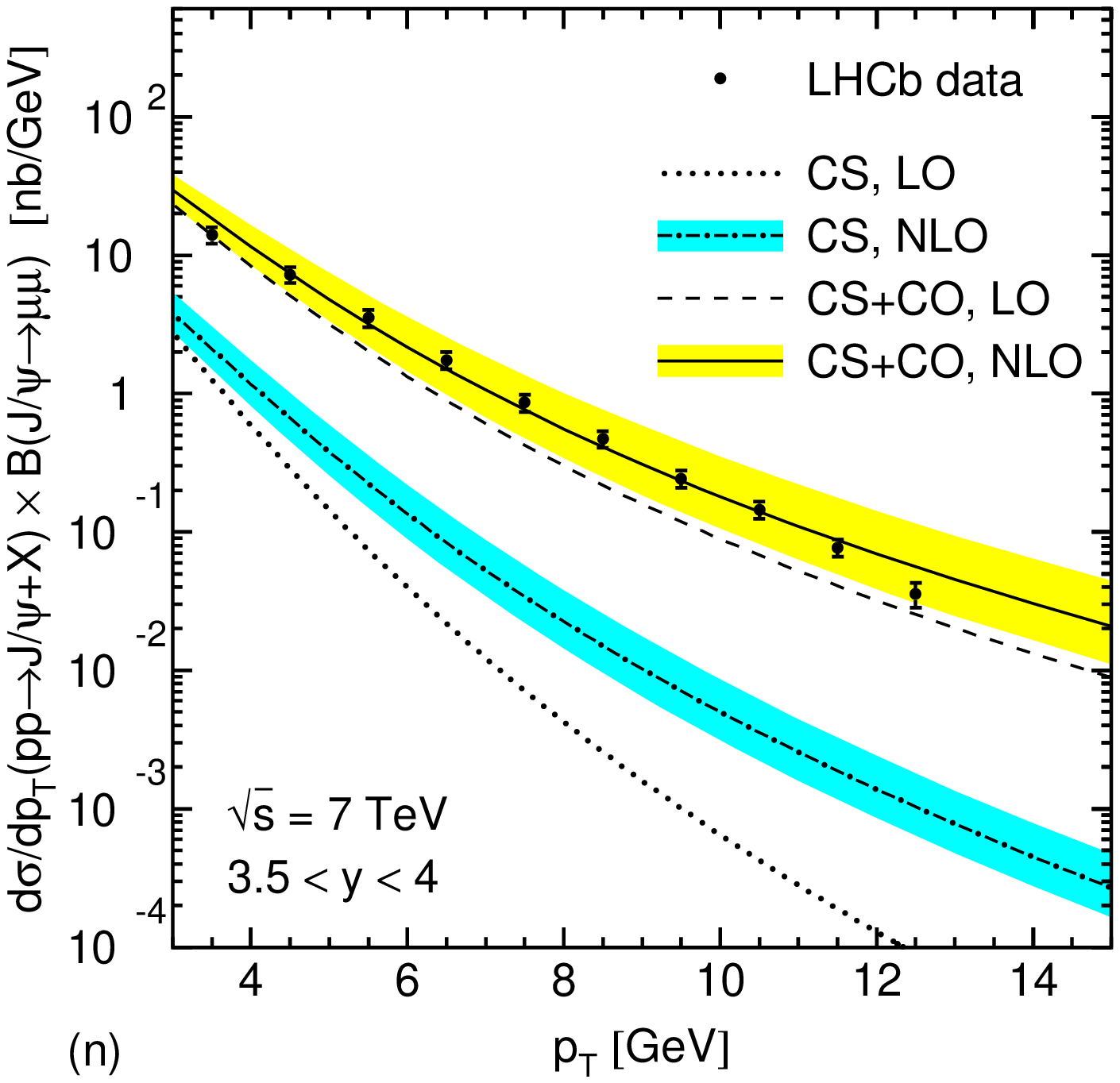}
\includegraphics[width=3.75cm]{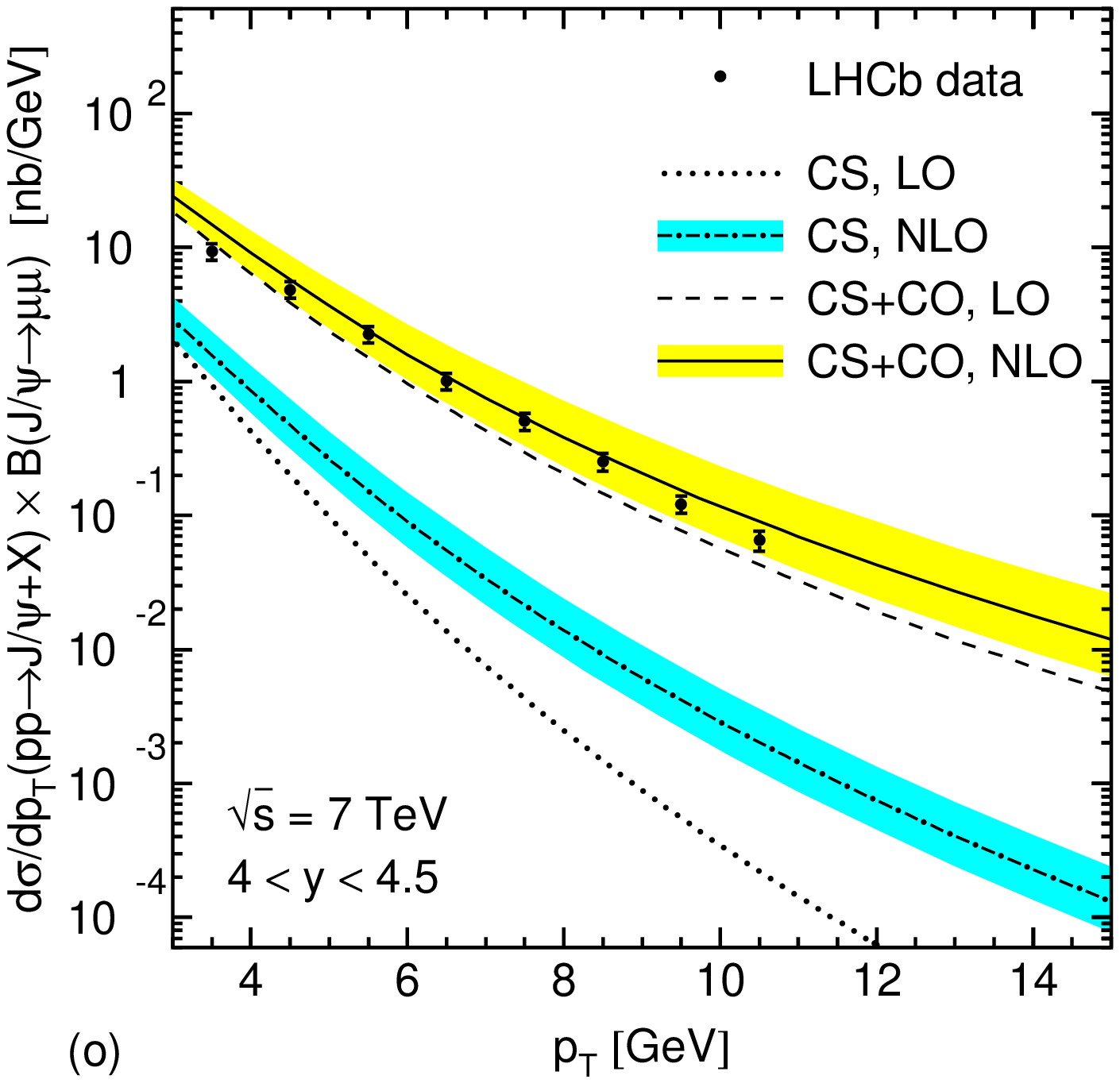}
\includegraphics[width=3.75cm]{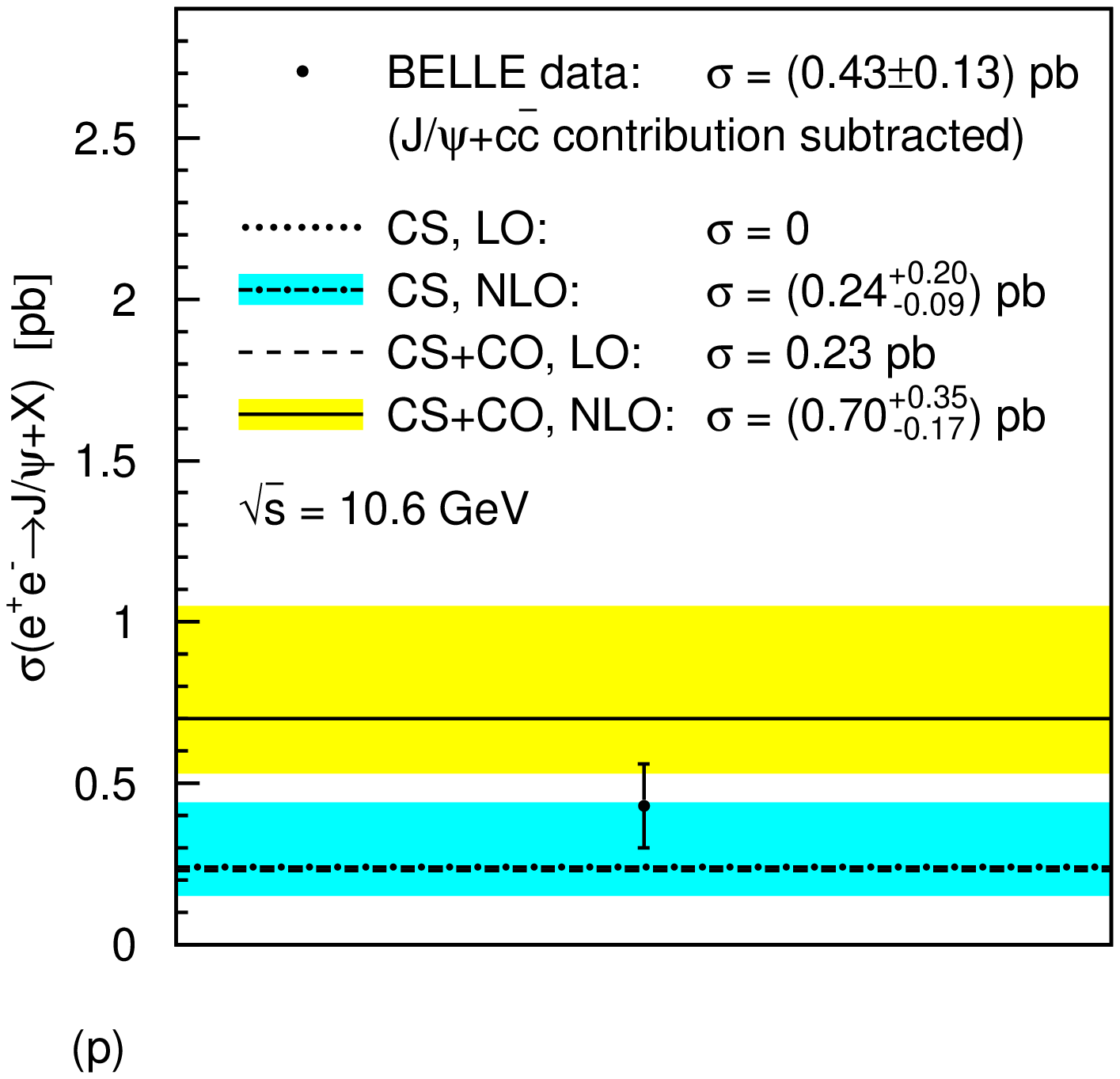}

\vspace{2pt}
\includegraphics[width=3.75cm]{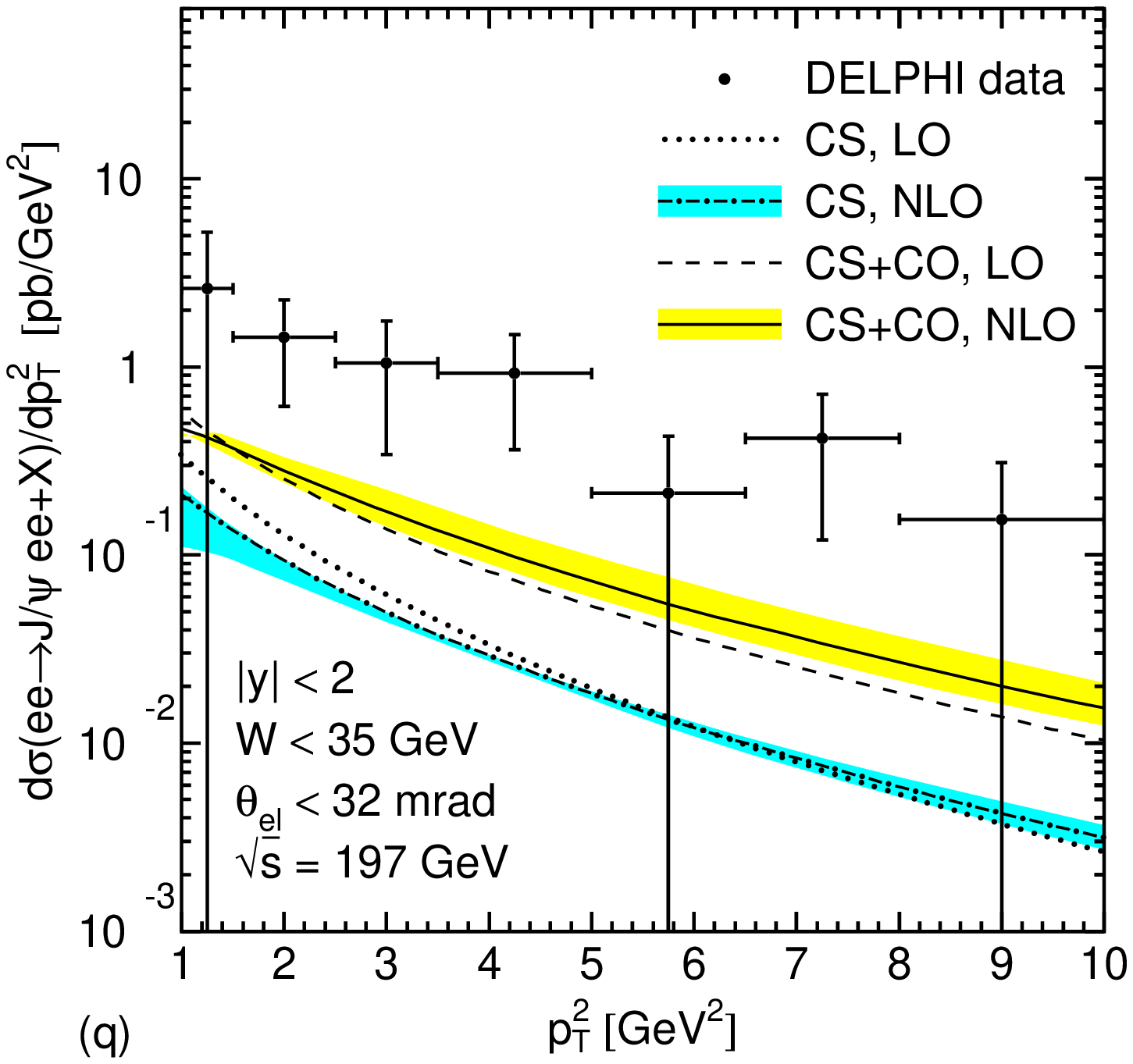}
\includegraphics[width=3.75cm]{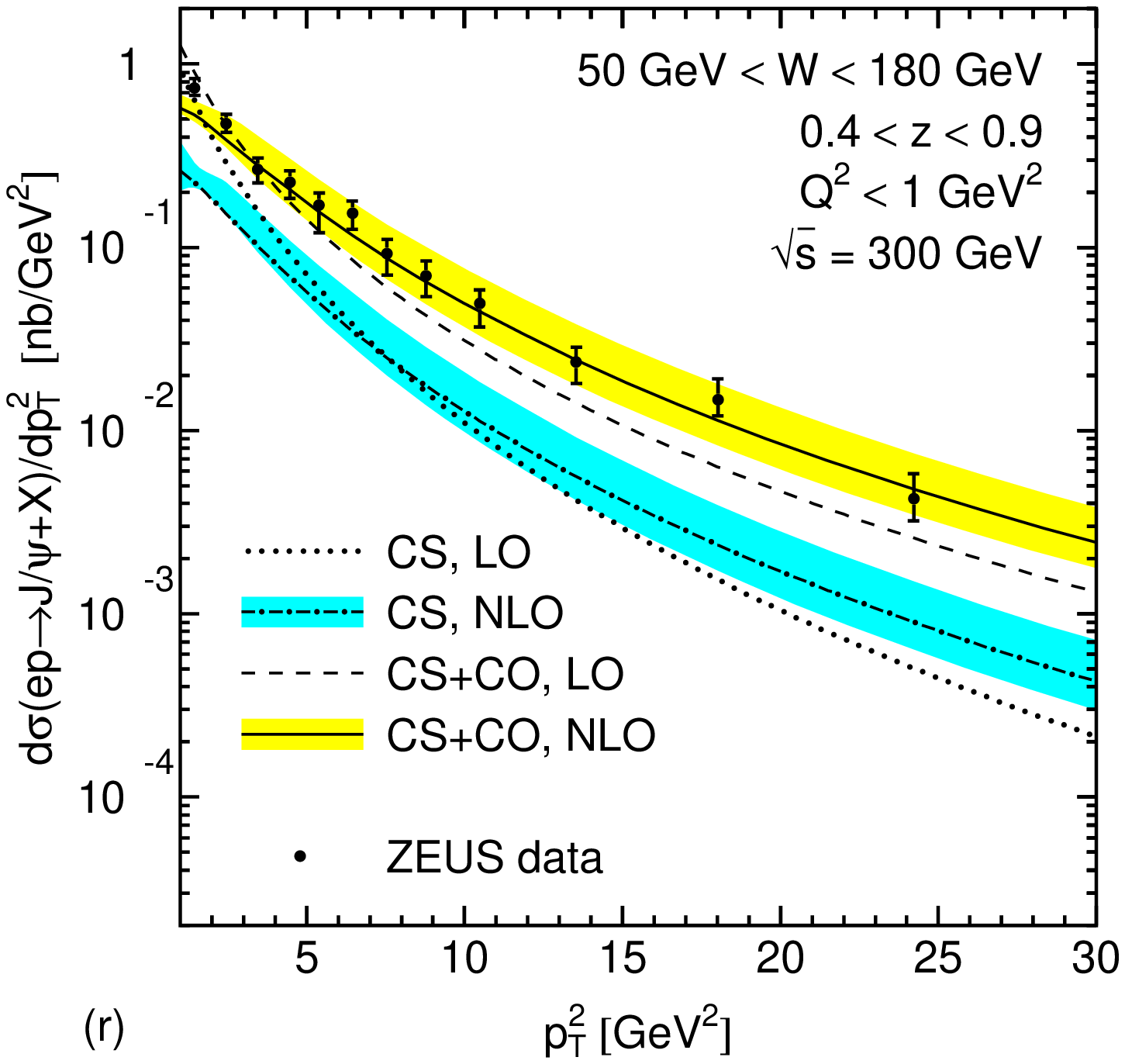}
\includegraphics[width=3.75cm]{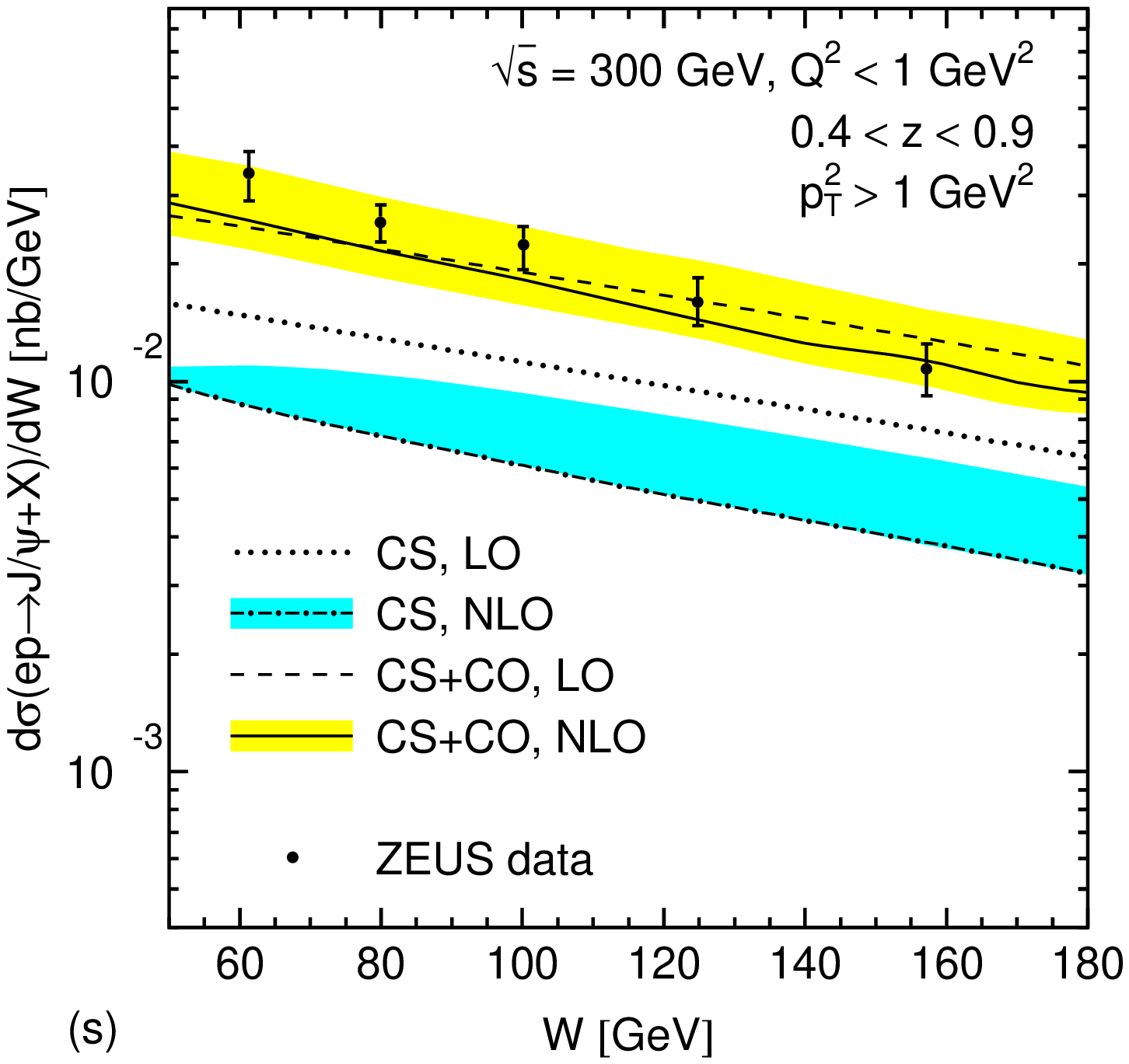}
\includegraphics[width=3.75cm]{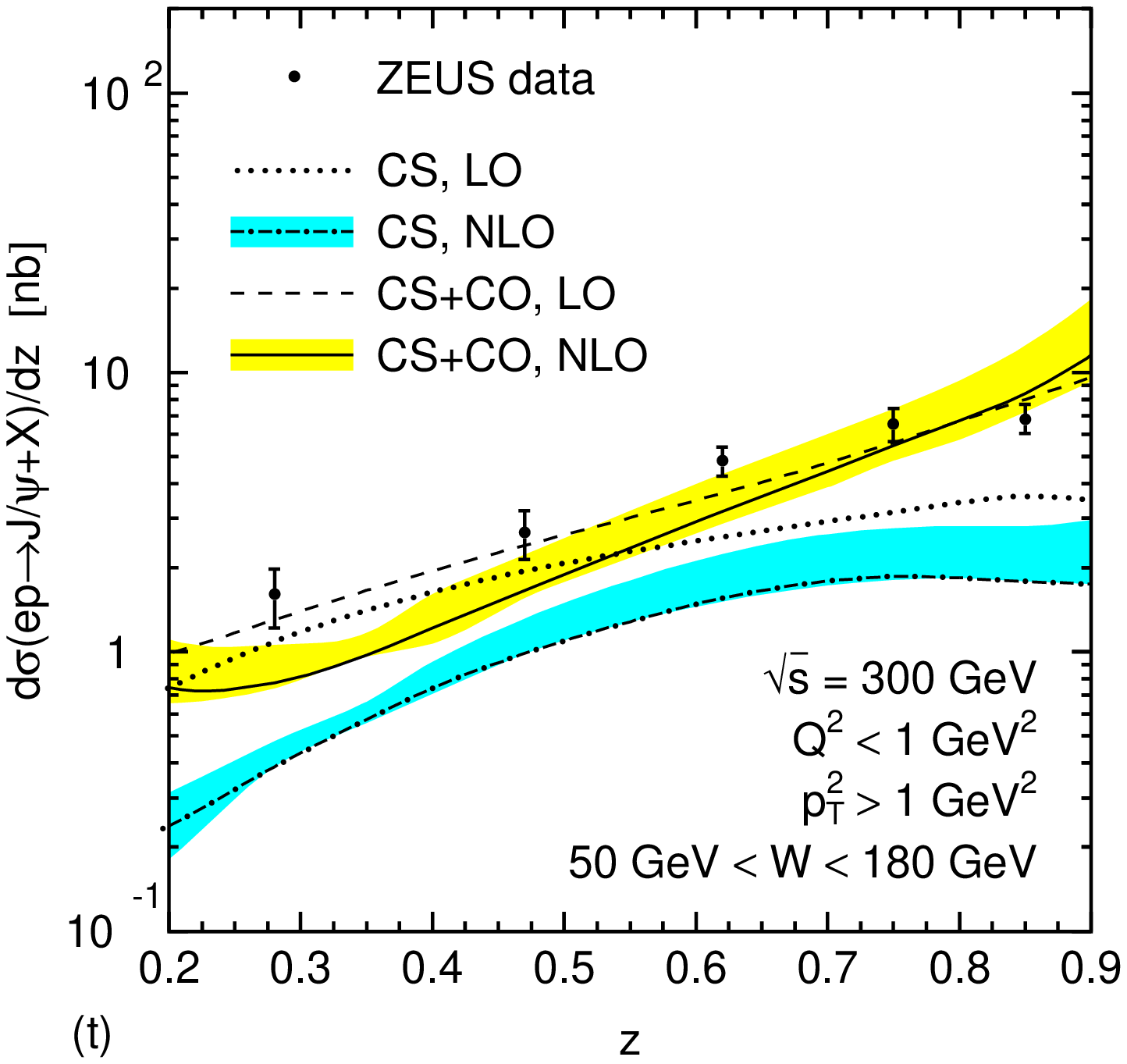}
\caption{\label{fig:fitgraphs1}
Plots a-t: Results of global fit \cite{Butenschoen:2011yh} compared to ALICE \cite{ALICEdata}, ATLAS \cite{ATLASdata}, BELLE \cite{:2009nj}, CDF \cite{Acosta:2004yw,Abe:1997jz}, CMS \cite{Khachatryan:2010yr}, DELPHI \cite{Abdallah:2003du}, LHCb \cite{Aaij:2011jh}, PHENIX \cite{Adare:2009js}, and ZEUS \cite{Chekanov:2002at} data. The blue bands are the color singlet model predictions, the yellow bands include the color octet contributions.}
\end{figure}
\addtocounter{figure}{-1}
\begin{figure}
\centering
\includegraphics[width=3.75cm]{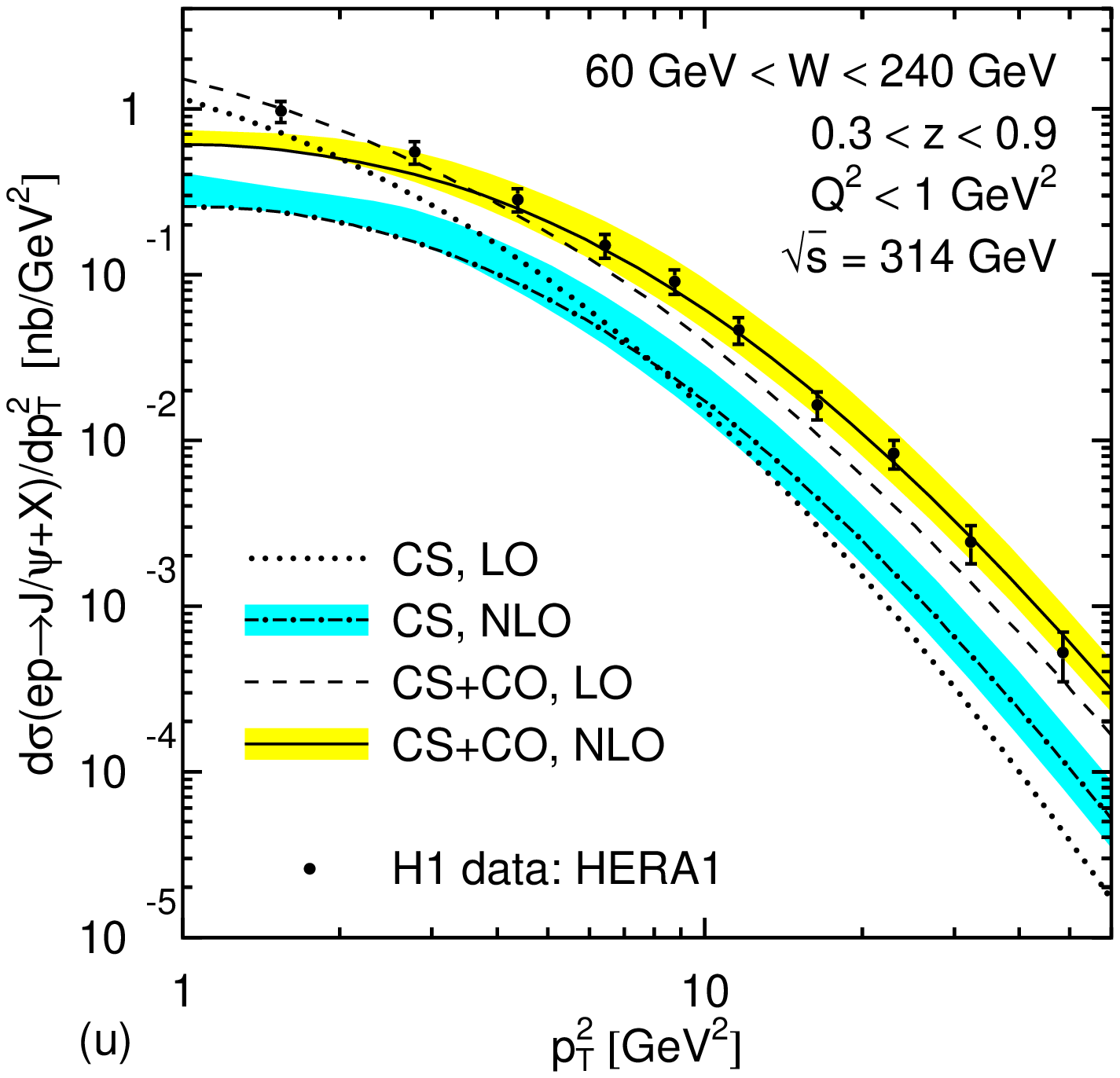}
\includegraphics[width=3.75cm]{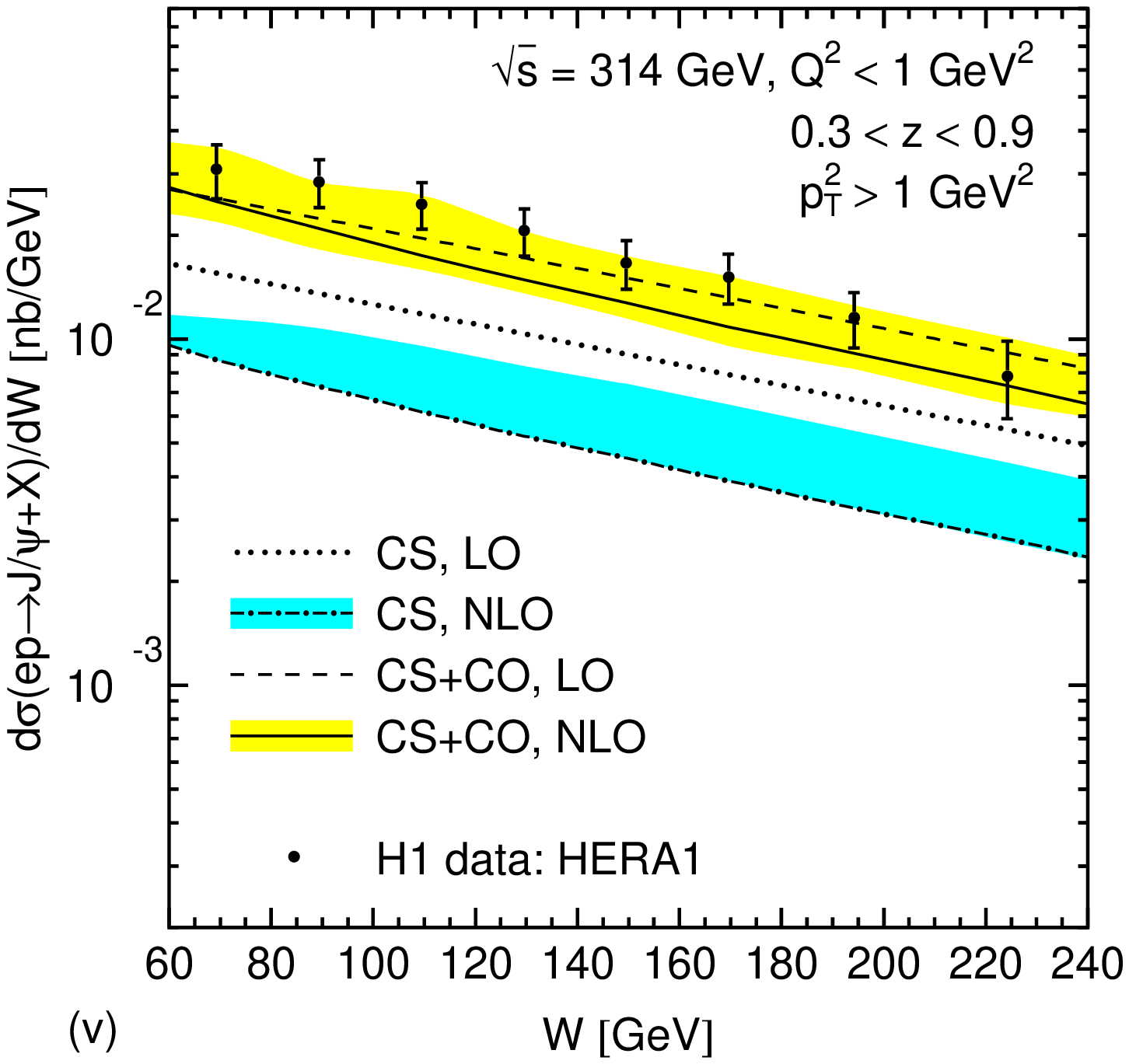}
\includegraphics[width=3.75cm]{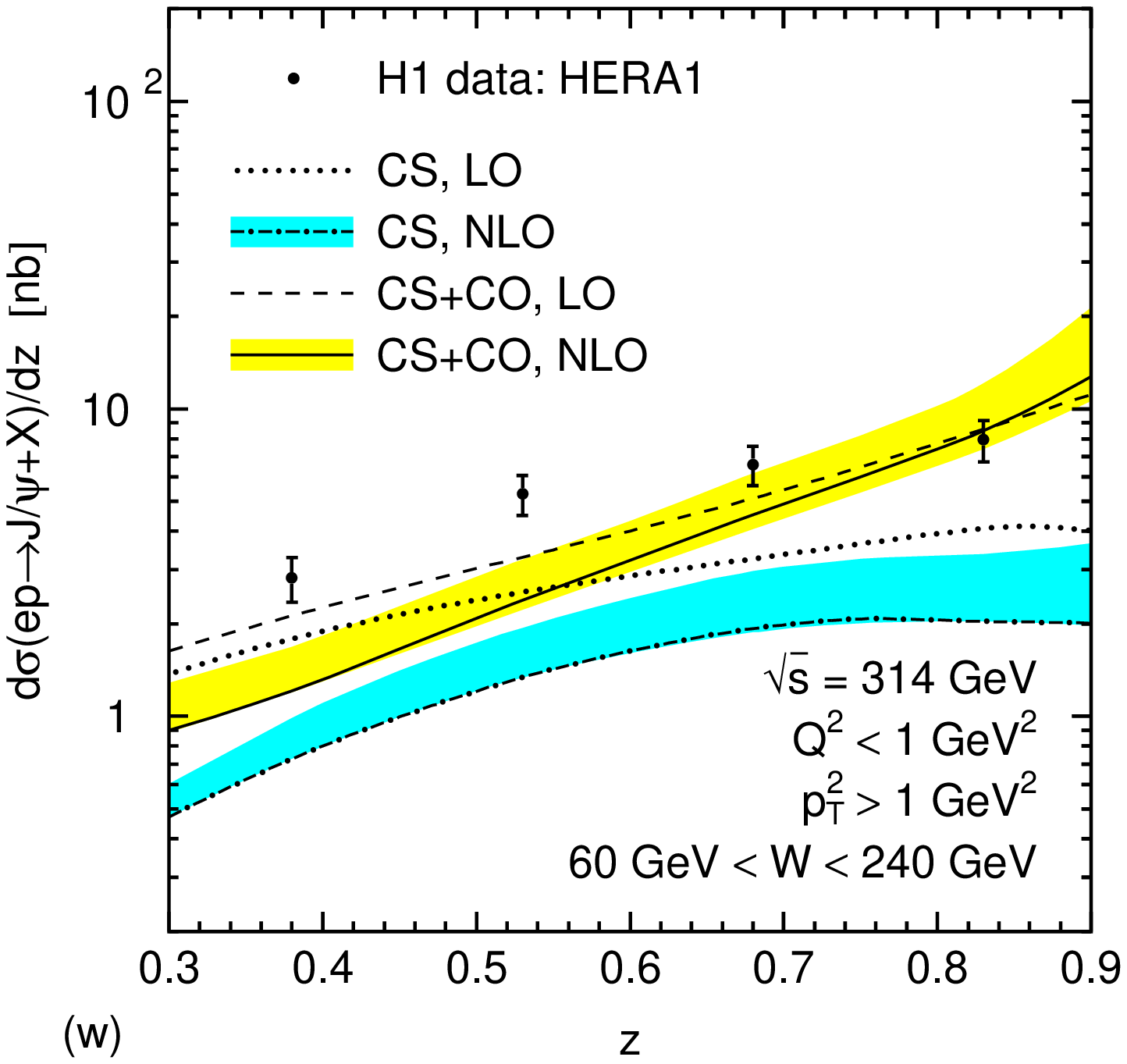}

\vspace{2pt}
\includegraphics[width=3.75cm]{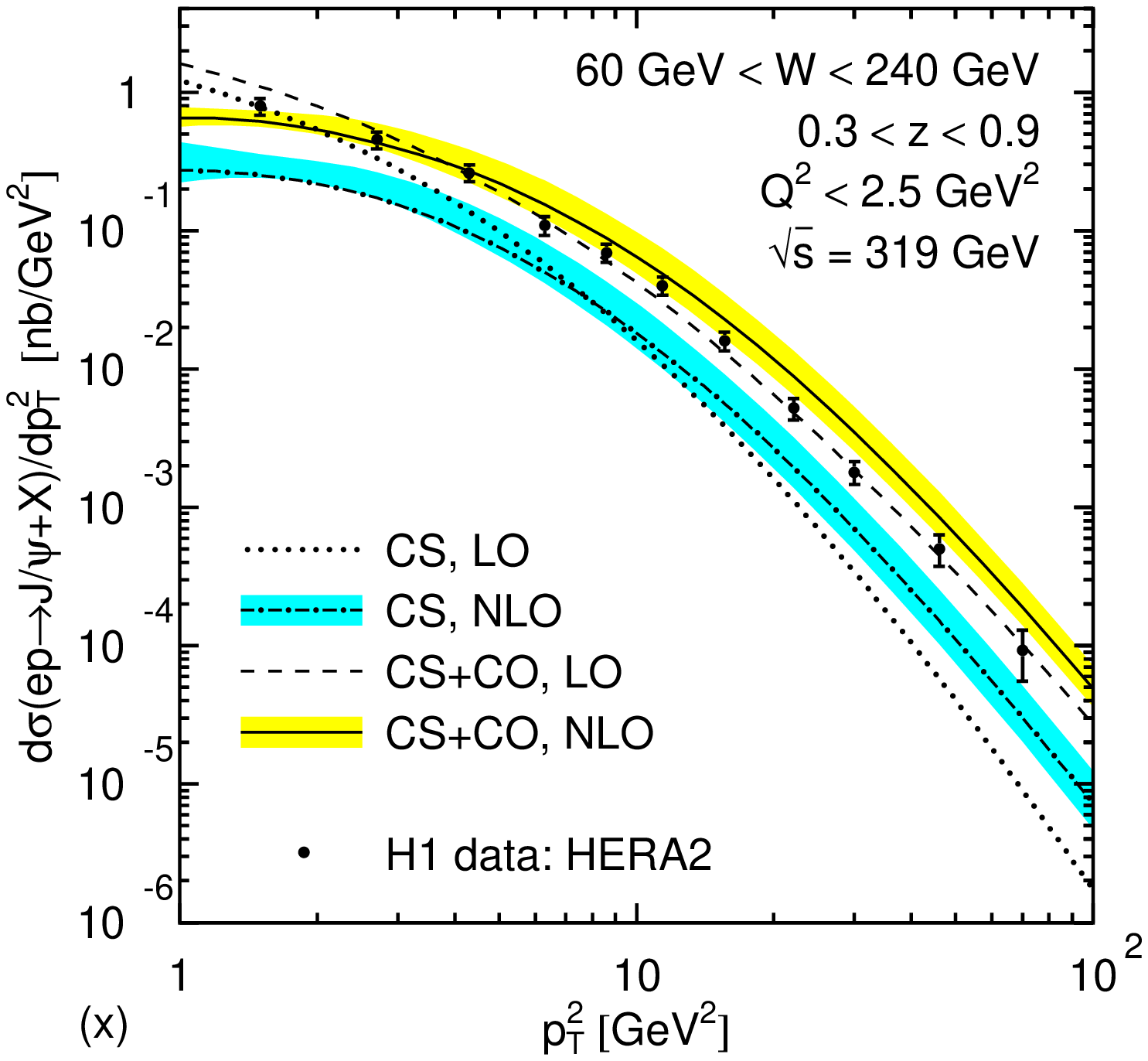}
\includegraphics[width=3.75cm]{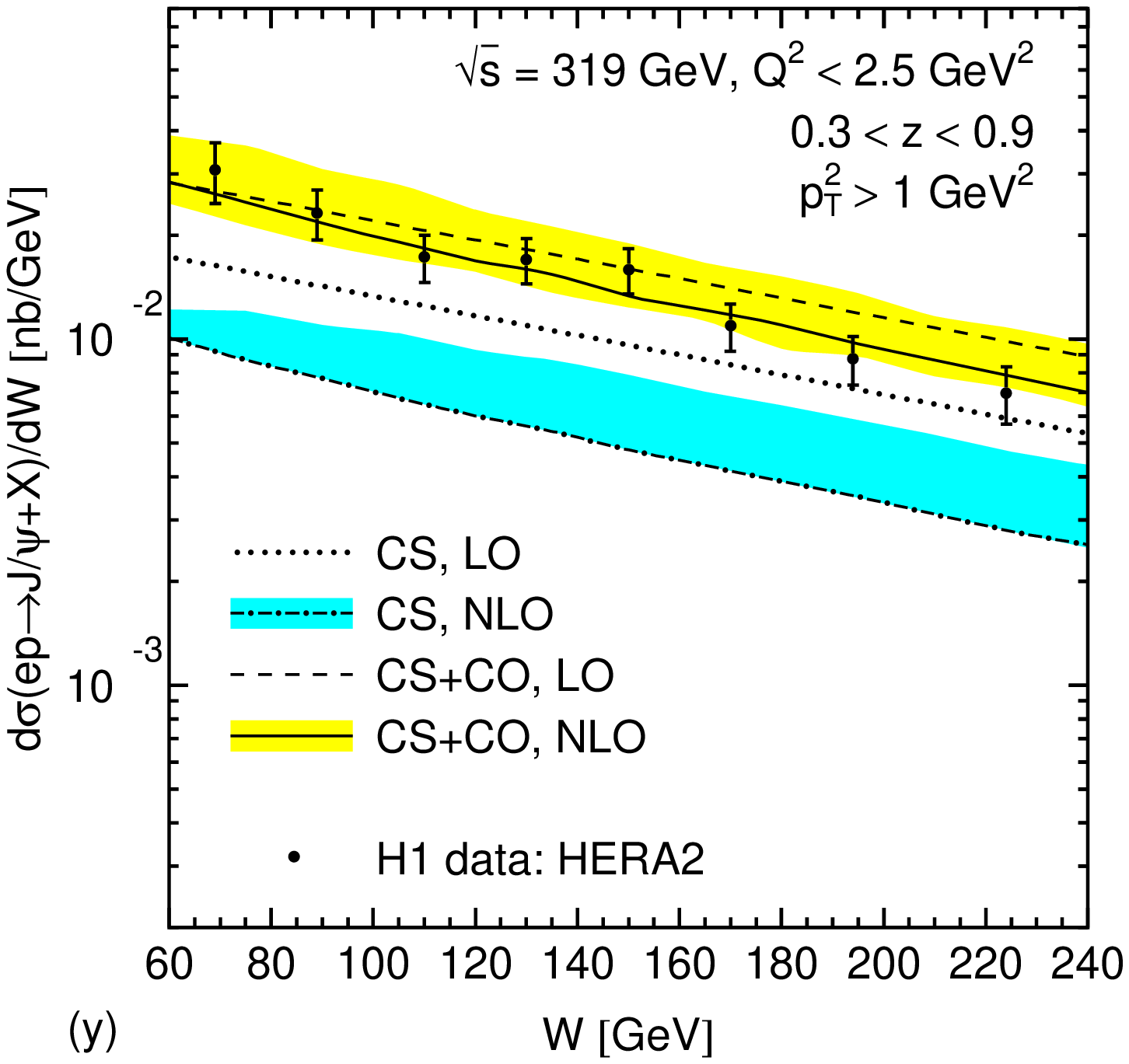}
\includegraphics[width=3.75cm]{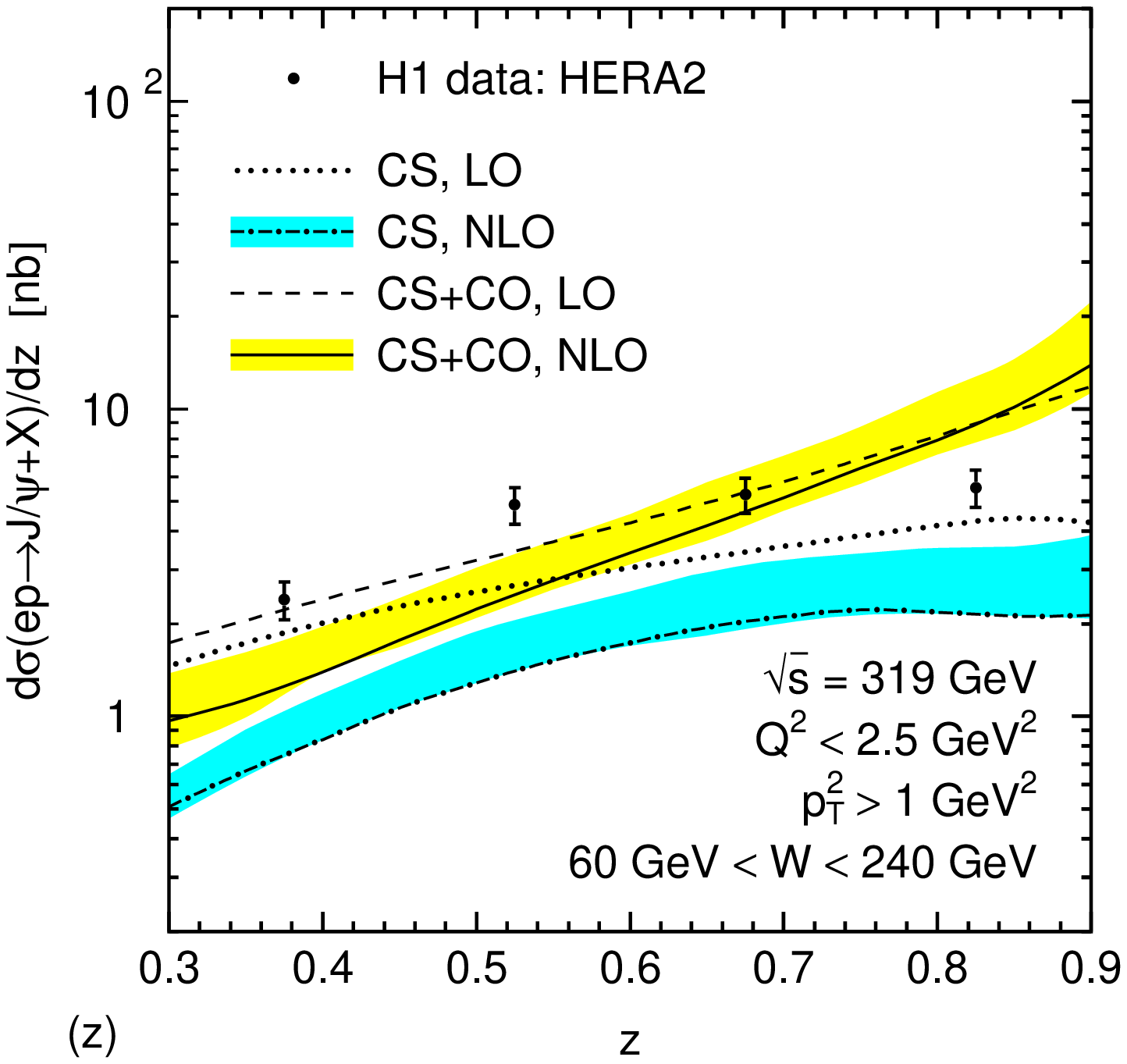}
\caption{Plots u-z (continuation): Results of global fit \cite{Butenschoen:2011yh} compared to H1 \cite{Adloff:2002ex,Aaron:2010gz} data. The blue bands are the color singlet model predictions, the yellow bands include the color octet contributions.}
\end{figure}

In \cite{Butenschoen:2011yh} a global NLO fit of the CO LDMEs to 194 data points of inclusive unpolarized $J/\psi$ production from 10 different experiments has been performed, see figure \ref{fig:fitgraphs1} and table \ref{tab:fit}. This extends the previous fit \cite{Butenschoen:2010rq} mentioned in the last section by including a lot more photoproduction and hadroproduction data and additionally including data from two-photon collisions measured by DELPHI \cite{Abdallah:2003du} and electron-positron collisions measured by BELLE \cite{:2009nj}. The new ingredients do not alter the fit values much, but the fit errors are strongly reduced. The reason is that in order to constrain the fit, input from basically just one photoproduction and one hadroproduction experiment is needed, and that input was already present in \cite{Butenschoen:2010rq}.

The global fit shows that at NLO all considered processes except perhaps the two-photon collisions can be described well when including the CO contributions. As explained in more detail in \cite{Butenschoen:2011yh}, the distribution in the inelasticity variable $z$ of photoproduction at HERA is now well described even at high $z$ (see figures \ref{fig:fitgraphs1}t, w and z), where the older Born analyses predicted a steep rise in the cross section not found in the data. The fact that the DELPHI data overshoots the NRQCD prediction is not worrying since the experimental errors are huge with just 16 events entering the data of figure \ref{fig:fitgraphs1}q. The CS contributions alone are on the other hand shown to fall clearly short of the data everywhere except for the BELLE total $e^+e^-$ cross section, see figure \ref{fig:fitgraphs1}p.

\begin{table}
\centering
\begin{tabular}{|c|c|c|}
\hline
 & Set A: Do not mind feed-downs & Set B: Subtract feed-downs first\\
\hline
$\langle {\cal O}^{J/\psi}(^1S_0^{[8]}) \rangle$ &
$(4.97\pm0.44)\times10^{-2}$~GeV$^3$ & $(3.04\pm0.35)\times10^{-2}$~GeV$^3$ \\
$\langle {\cal O}^{J/\psi}(^3S_1^{[8]}) \rangle$ &
$(2.24\pm0.59)\times10^{-3}$~GeV$^3$ & $(1.68\pm0.46)\times10^{-3}$~GeV$^3$ \\
$\langle {\cal O}^{J/\psi}(^3P_0^{[8]}) \rangle$ &
$(-1.61\pm0.20)\times10^{-2}$~GeV$^5$ & $(-9.08\pm1.61)\times10^{-3}$~GeV$^5$ \\
\hline
\end{tabular}
\caption{\label{tab:fit} Results of global fit \cite{Butenschoen:2011yh} for the $J/\psi$ CO LDMEs. Set A corresponds to the main fit shown in figure \ref{fig:fitgraphs1}. In set B, estimated feed-down contributions from higher charmonium states were subtracted from the prompt data prior to fitting (hadroproduction: 36\%, photoproduction: 15\%, $\gamma\gamma$ scattering: 9\%, $e^+e^-$ annihilation: 26\%).}
\end{table}

\subsection{Polarization observables}

\begin{figure}
\begin{tabular}{|c|c|c|}
\hline
\includegraphics[width=4.63cm]{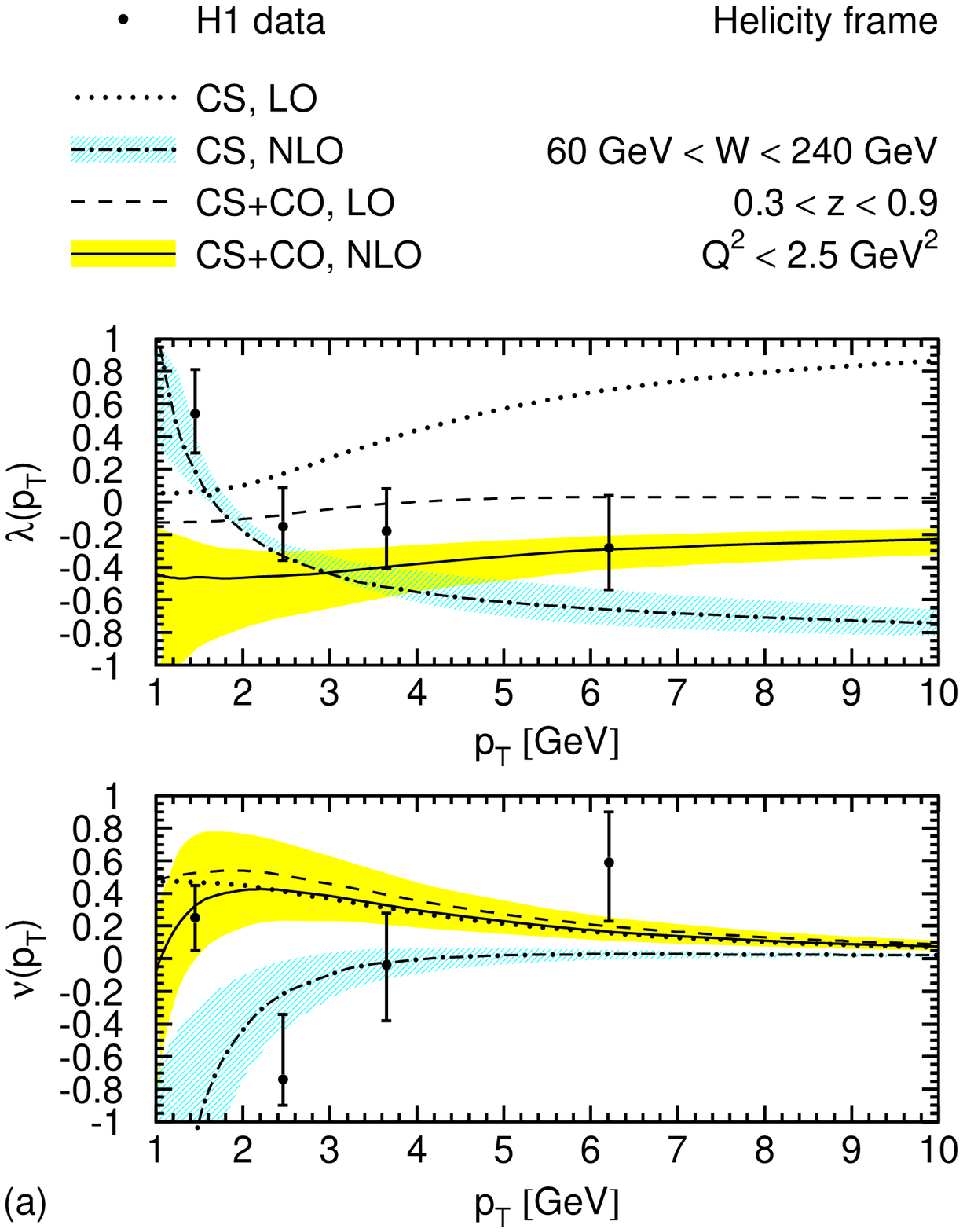}
&
\includegraphics[width=4.63cm]{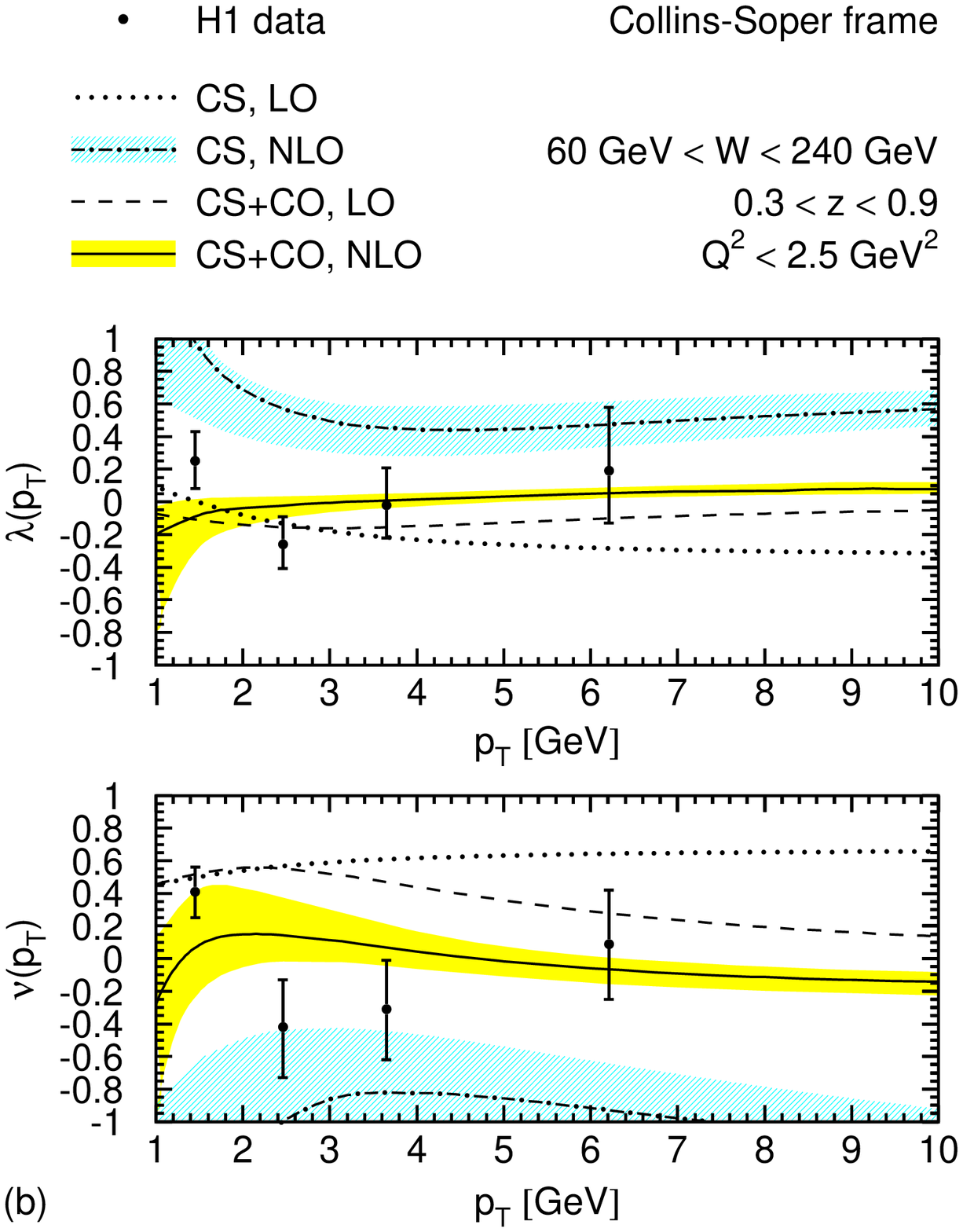}
&
\includegraphics[width=4.63cm]{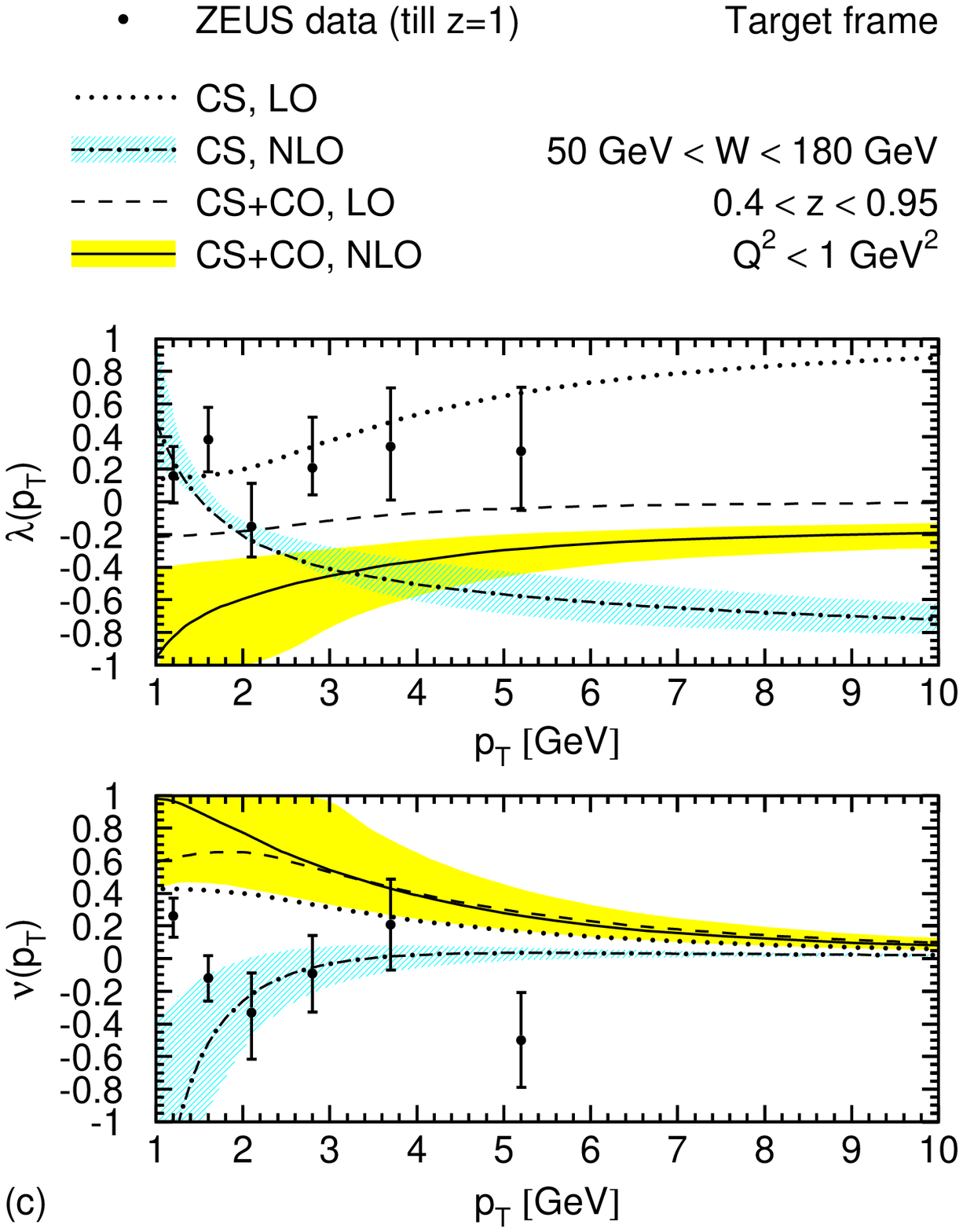}
\\ \hline
\includegraphics[width=4.63cm]{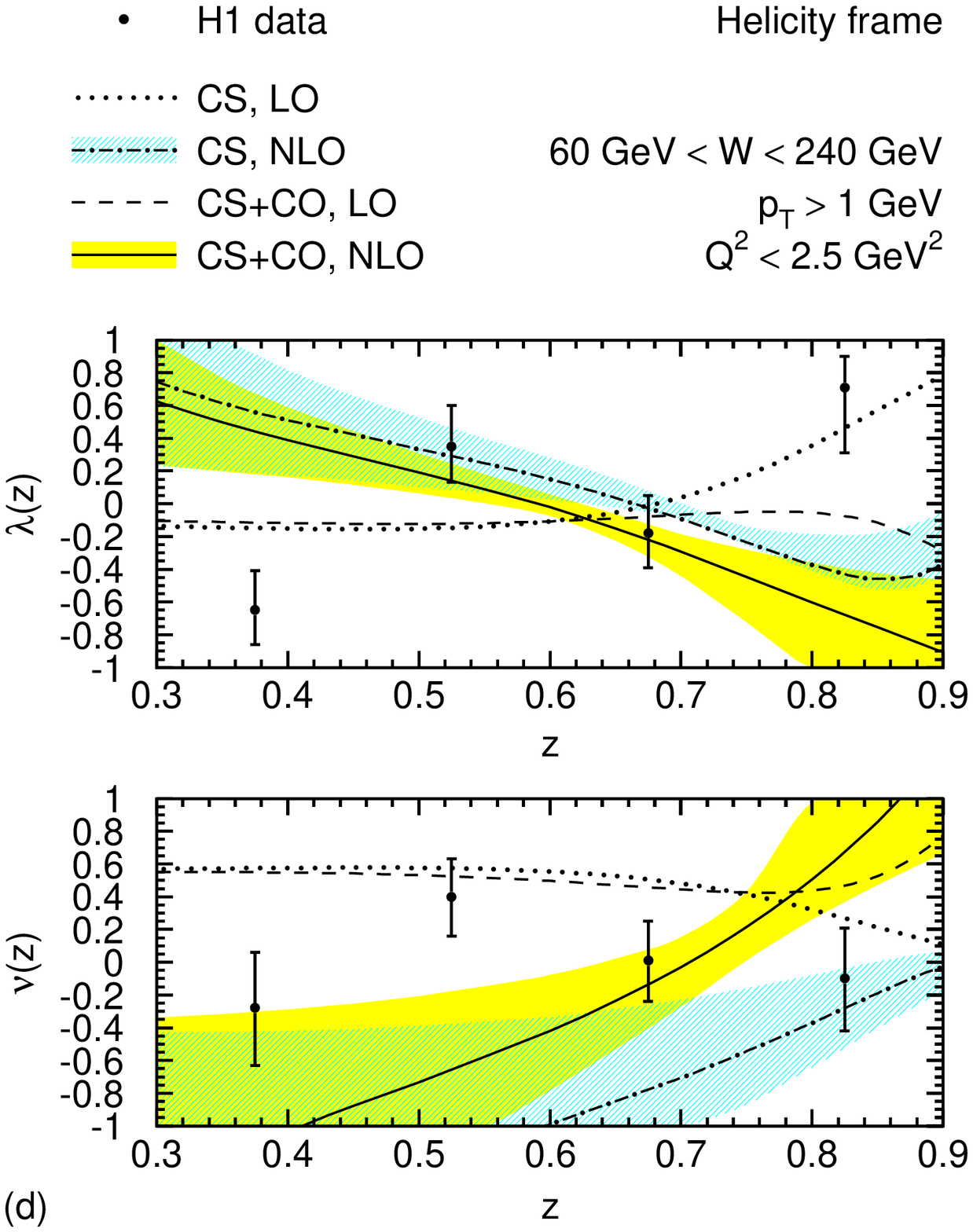}
&
\includegraphics[width=4.63cm]{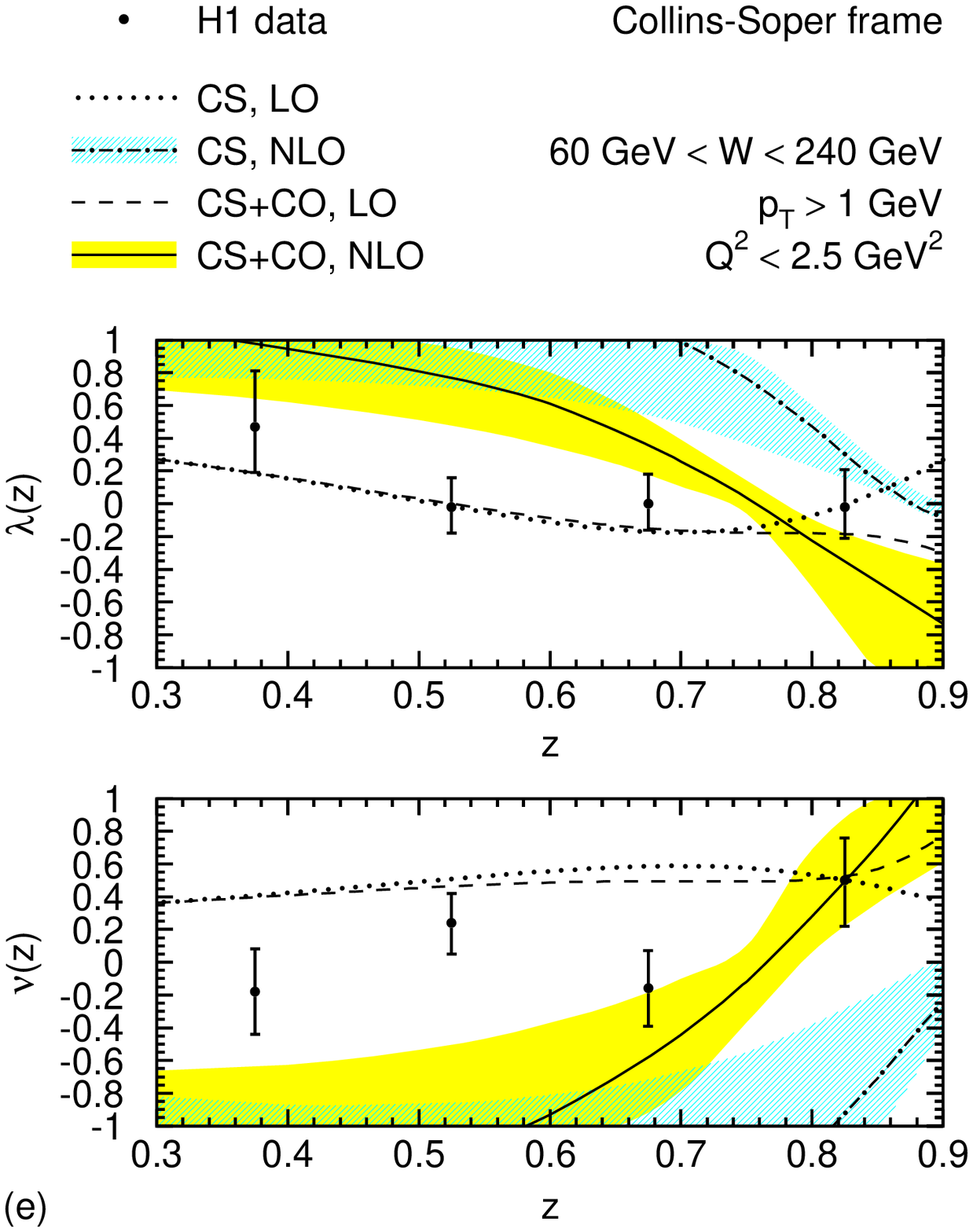}
&
\includegraphics[width=4.63cm]{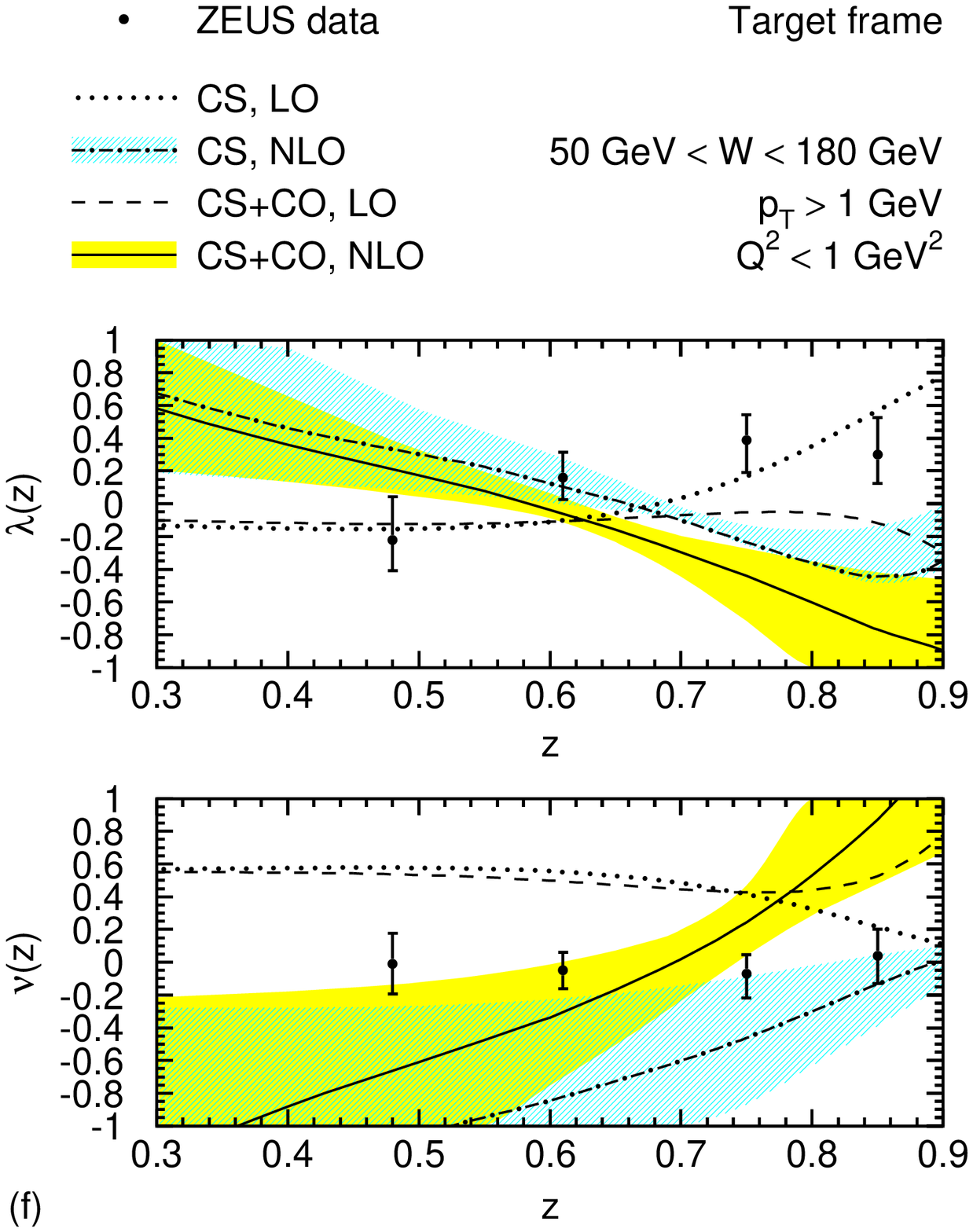}
\\ \hline
\end{tabular}
\caption{\label{fig:PhotoPol} Polarization parameters $\lambda$ and $\nu$ for direct photoproduction at HERA using CO LDME set B of table \ref{tab:fit}, compared to H1 \cite{Aaron:2010gz} and ZEUS \cite{:2009br} data. Blue bands: Uncertainties of NLO CS curve due to scale variations. Yellow bands: Uncertainties of NLO CS+CO curve due to scale variations and uncertainties of the CO LDMEs. From \cite{PhotoPolLetter}.}
\end{figure}

\begin{figure}
\centering
\includegraphics[height=4cm]{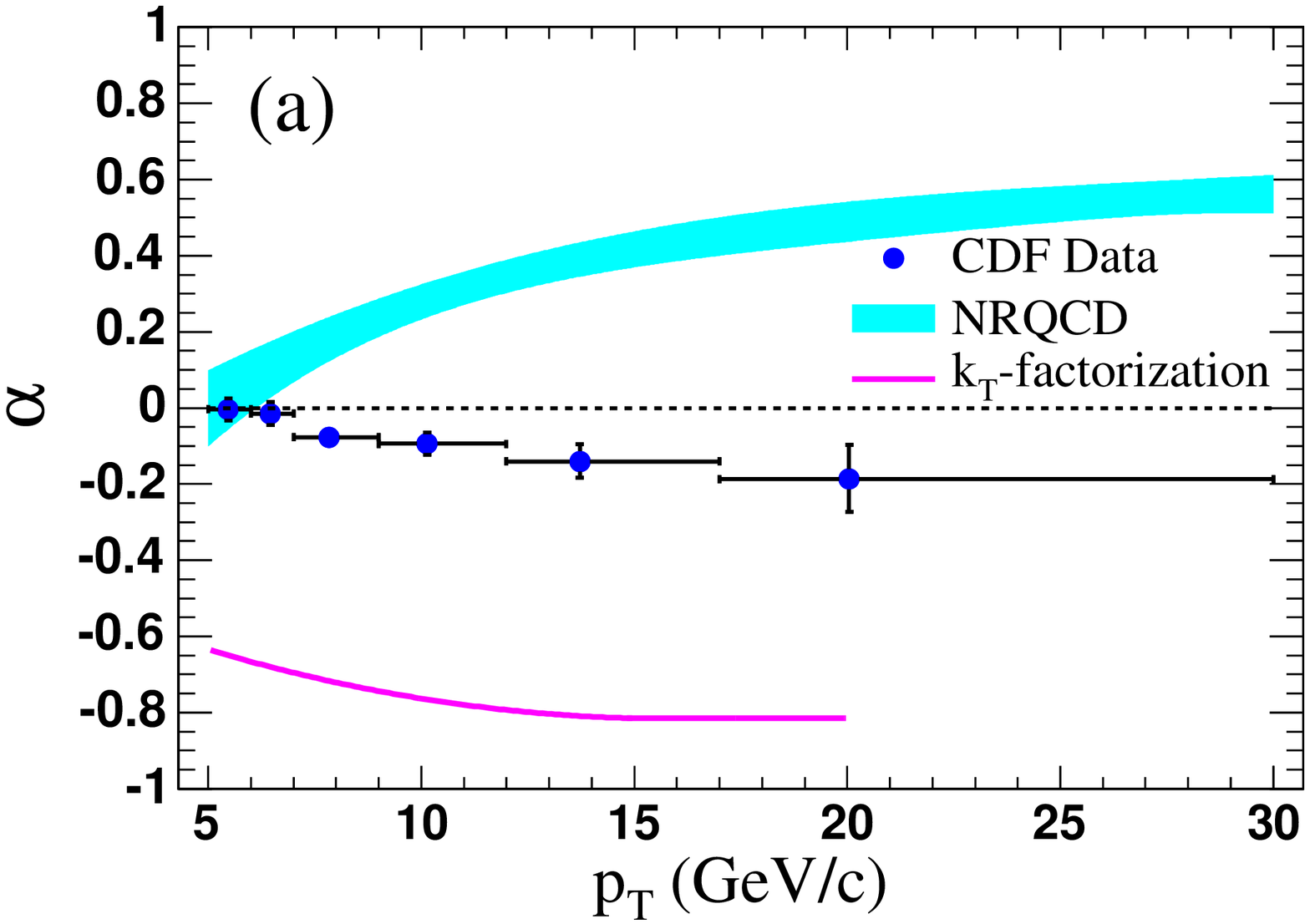}
\hspace{1cm}
\includegraphics[height=4cm]{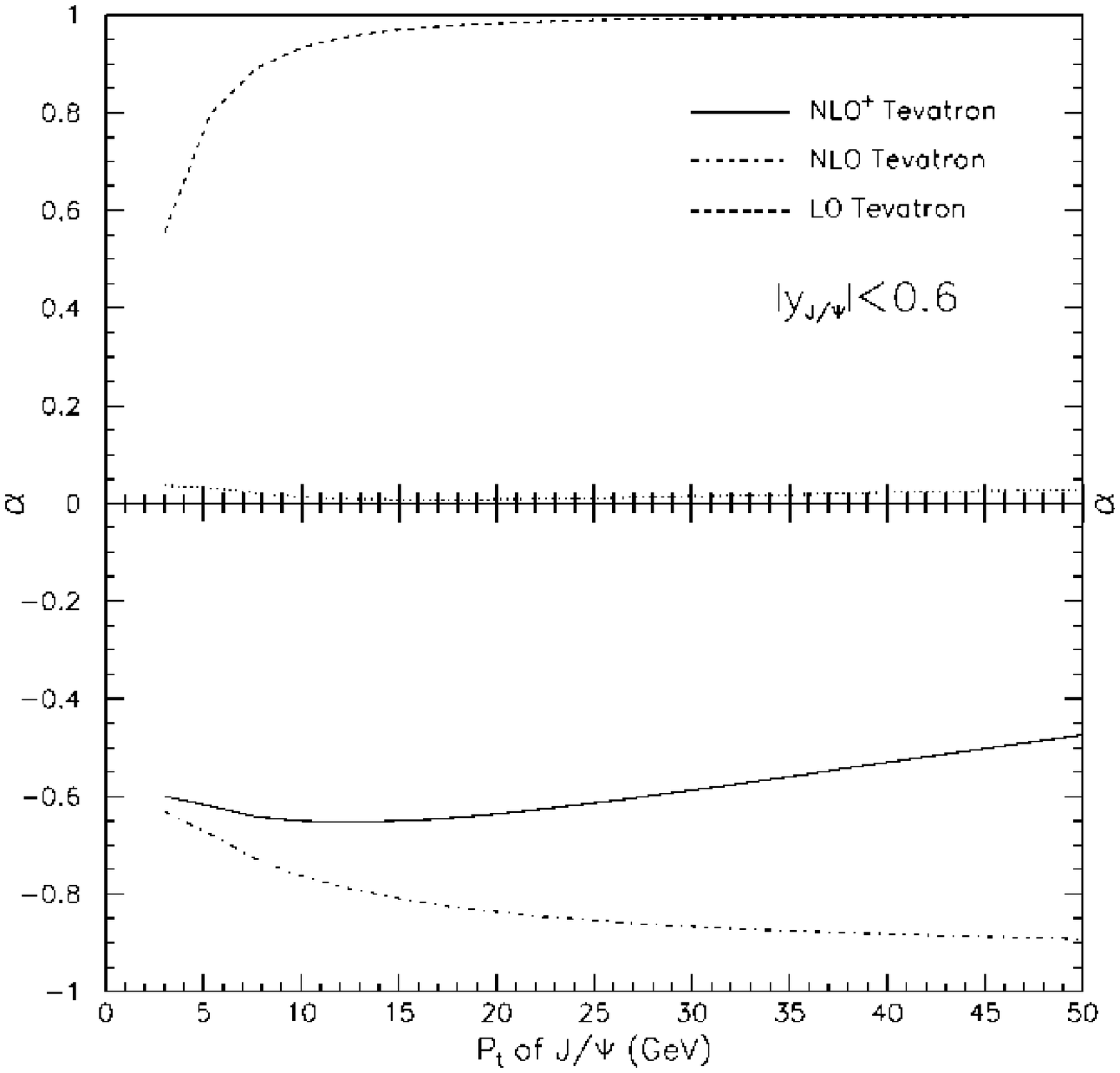}
\caption{\label{fig:Tevpol} Polarization parameter $\alpha$ in the helicity frame at the Tevatron \cite{Abulencia:2007us}, compared to a LO NRQCD prediction and the CSM $k_T$ factorization prediction \cite{Baranov:2002cf} (left) and the LO and NLO CSM prediction in collinear factorization \cite{Gong:2008sn} (right). From \cite{Abulencia:2007us} (left) and \cite{Gong:2008sn} (right).}
\end{figure}

Polarized NLO $J/\psi$ production cross sections have been evaluated within the CSM, for direct photoproduction \cite{Artoisenet:2009xh} and hadroproduction \cite{Gong:2008sn}, and for the $^1S_0^{[8]}$ and $^3S_1^{[8]}$ intermediate states in hadroproduction \cite{Gong:2008ft}. Recently, for the first time, polarized NLO cross sections including all CO contributions have been calculated, namely for direct photoproduction at HERA \cite{PhotoPolLetter}. In figure \ref{fig:PhotoPol}, the predictions for the polarization parameters $\lambda$ and $\nu$ are compared to data. They are defined by the angular momentum distribution of the decay muons via
\begin{equation}
\frac{d\sigma(J/\psi\to\mu^+\mu^-)}{d\cos\theta\, d\phi} \propto 1 + \lambda \cos^2 \theta + \mu \sin 2\theta \cos \phi  + \frac{\nu}{2} \sin^2\theta \cos 2\phi, \label{eq:PolParams}
\end{equation}
where $\theta$ and $\phi$ are the polar and azimuthal angles of the $\mu^+$ in the $J/\psi$ rest frame for specific choices of the coordinate axes. $\lambda=0$ corresponds to unpolarized $J/\psi$, whereas $\lambda=+1$ (-1) stands for fully transversely (longitudinally) polarized $J/\psi$. Unfortunately, the H1 \cite{Aaron:2010gz} and ZEUS \cite{:2009br} data do not yet allow to distinguish the production mechanisms clearly, but kinematical regions can be identified, in which a distinction could be possible in a future more precise $ep$ collider experiment: At higher $p_T$, NRQCD predicts the $J/\psi$ to be largely unpolarized in contrast to the CSM predictions. In the $z$ distributions, however, the scale uncertainties are sizeable and the error bands of the CSM and NRQCD largely overlap. The LO calculation corresponding to that NLO analysis has first been performed in \cite{Beneke:1998re}.

As for hadroproduction at the Tevatron, the CDF \cite{Abulencia:2007us} measurement shows that the $J/\psi$'s are largely unpolarized, whereas the NLO CSM calculation \cite{Gong:2008sn} predicts largely longitudinally polarized $J/\psi$, see figure \ref{fig:Tevpol}. The parameter $\alpha$ equals $\lambda$ in the definition (\ref{eq:PolParams}). Predictions including all the CO contributions have so far only been made at LO \cite{Cho:1994ih}.

\subsection{Improving the Color Singlet Model: k$_{\mathbf{T}}$ factorization}

In heavy quarkonium production, the hard scattering scales are typically much lower than the collision energies, and the tested longitudinal momentum fractions $x$ of the partons inside the protons are so small that the partons' transverse momenta $k_T$ are of the same order as the longitudinal momenta and should hence not be neglected. That is the basic idea behind using the $k_T$ factorization approach \cite{Gribov:1984tu} in quarkonium production calculations. The initial gluons are therefore off-shell in this formalism. The partonic cross sections, which are so far only evaluated at LO in $\alpha_s$, are then convoluted with unintegrated, $k_T$ dependent gluon parton distribution functions (PDFs), which are derived from the usual gluon PDFs either in a DGLAP \cite{Gribov:1972ri}, a BFKL \cite{Kuraev:1977fs} or a so called CCFM \cite{Ciafaloni:1987ur} approach. This derivation of the unintegrated PDFs is certainly the most subtle point here. Usually, only CS contributions are considered. The $k_T$ factorization method gives very good descriptions of the $J/\psi$ photo- and electroproduction at HERA \cite{Lipatov:2002tc,Kniehl:2006sk}, and has also been applied for hadroproduction of $J/\psi$, $\chi_c$ and $\Upsilon$ at Tevatron \cite{Baranov:2002cf,Kniehl:2006sk,Kniehl:2006vm} and RHIC \cite{Baranov:2007dw}. The Monte Carlo program CASCADE \cite{Jung:2000hk} also successfully simulates initial gluon radiation within the $k_T$ factorization approach applying the CCFM \cite{Ciafaloni:1987ur} evolution equation.

\subsection{Improving the Color Singlet Model: ``NNLO$^\ast$''}

The CSM could describe the hadroproduction data better if the next-to-next-to-leading-order (NNLO) corrections had a large $K$ factor like the NLO corrections. NNLO corrections consist of three parts: Two-loop contributions, one-loop times tree-level contributions and pure real corrections. Only the sum of the three parts is infrared finite and gives the physical result. Unfortunately, to date only the real corrections are calculated. In \cite{Artoisenet:2008fc}, a ``NNLO$^\ast$'' correction was defined. It consists only of the real corrections, which are made finite by cutting off phase space parts around the singularities in which $k_i\cdot k_j<x_{\mathrm{cut}}$, with $k_i$ and $k_j$ being momenta of external light QCD partons. The ``NNLO$^\ast$'' band is computed by shifting this cutoff parameter $x_\mathrm{cut}$. This band reminds us that the NNLO cross sections can be expected to have a flatter $p_T$ dependence than the NLO ones, and it is possible that the NNLO corrections may indeed be large and positive, like the ''NNLO$^\ast$`` ones.

\section{Open heavy flavour production}

\subsection{Theory frameworks}

Heavy flavoured hadrons are hadrons consisting of one heavy quark and one or two light quarks. Examples are the $D$ mesons (charm plus one light quark) and the $B$ mesons (bottom plus one light quark). The production of these particles is described by the fragmentation of outgoing heavy or light QCD partons into the heavy flavoured hadrons. The partonic cross sections are thus folded not only with the PDFs but also with nonperturbative fragmentation functions (FFs), whose exact definition and theoretical interpretation depend on the calculational scheme used. There are two main traditional schemes, which are valid in complementary kinematical regions: The fixed-flavour-number scheme (FFNS), which was also applied in all the heavy quarkonium calculations of section \ref{sec:heavyquarkonium}, and the zero-mass variable-flavour-number scheme (ZM-VFNS). Let us for simplicity assume charm $c$ as the heavy quark. In the FFNS, we then have only the light quarks $u,d,s$ and gluons as incoming particles, the heavy quark $c$ only appears as a final state particle, and the heavy quark mass $m_c$ is kept finite. Here, we have two kinematical scales: $m_c$ and the typical scale $Q$, which could be the hadron's transverse momentum. This scheme is valid only at $m_c^2\lessapprox Q^2$, because at very large $Q^2$, large logarithms $\log(Q^2/m_c^2)$ spoil the convergence of the perturbative expansion. On the other hand, the ZM-VFNS is the classic parton model. Here, the charm is treated massless like a light quark, and it appears both as an incoming and an outgoing parton. Instead of the quasi-collinear logarithms $\log(Q^2/m_c^2)$, genuinely collinear divergent terms appear, which are factorized into the charm quark PDFs and FFs. Since any $m_c$ dependent terms are missing, the ZM-VFNS is a good approximation only in the limit $m_c^2\ll Q^2$. Although the FFNS alone can already describe the data well in the currently accessed kinematical regions, one would like to have a combined scheme, which interpolates between the FFNS and the ZM-VFNS, and is by itself valid at all scales~$Q^2$. There are currently two of these interpolating schemes on the market: The general-mass variable-flavour-number scheme (GM-VFNS) \cite{Kniehl:2004fy} and the fixed-order NLL scheme (FONLL) \cite{Cacciari:1998it}.

The GM-VFNS is an extension of the ZM-VFNS in such a way that in Feynman diagrams where $c$ appears only as an outgoing parton, we do now consider a nonzero heavy quark mass $m_c$, while when it does also/only appear as an incoming particle, the heavy quark mass is still kept zero like in the original ZM-VFNS. The bulk of the heavy mass dependence is now taken into account, and the applicability of the ZM-VFNS is lowered down to scales of about a few times the heavy quark mass. The large $\log(Q^2/m_c^2)$ terms now appearing are factorized into the heavy quark PDFs and FFs and resummed using the DGLAP \cite{Gribov:1972ri} evolution equation according to the QCD factorization theorems, which are proven to hold also in the case of these quasi-collinear logarithms \cite{Collins:1998rz}.

In the FONLL scheme the predictions of the FFNS and the ZM-VFNS are overlaid by using a $Q=p_T$ dependent weight function, such that the FFNS and ZM-VFNS are recovered in the respective $p_T$ limits. Additionally, the heavy quark FFs contain perturbative pieces at the starting scale $\mu_0=m_c$, such that the ZM-VFNS result matches the FFNS one at NLO.

\subsection{Applications}

\begin{figure}
\centering
\includegraphics[height=4.4cm]{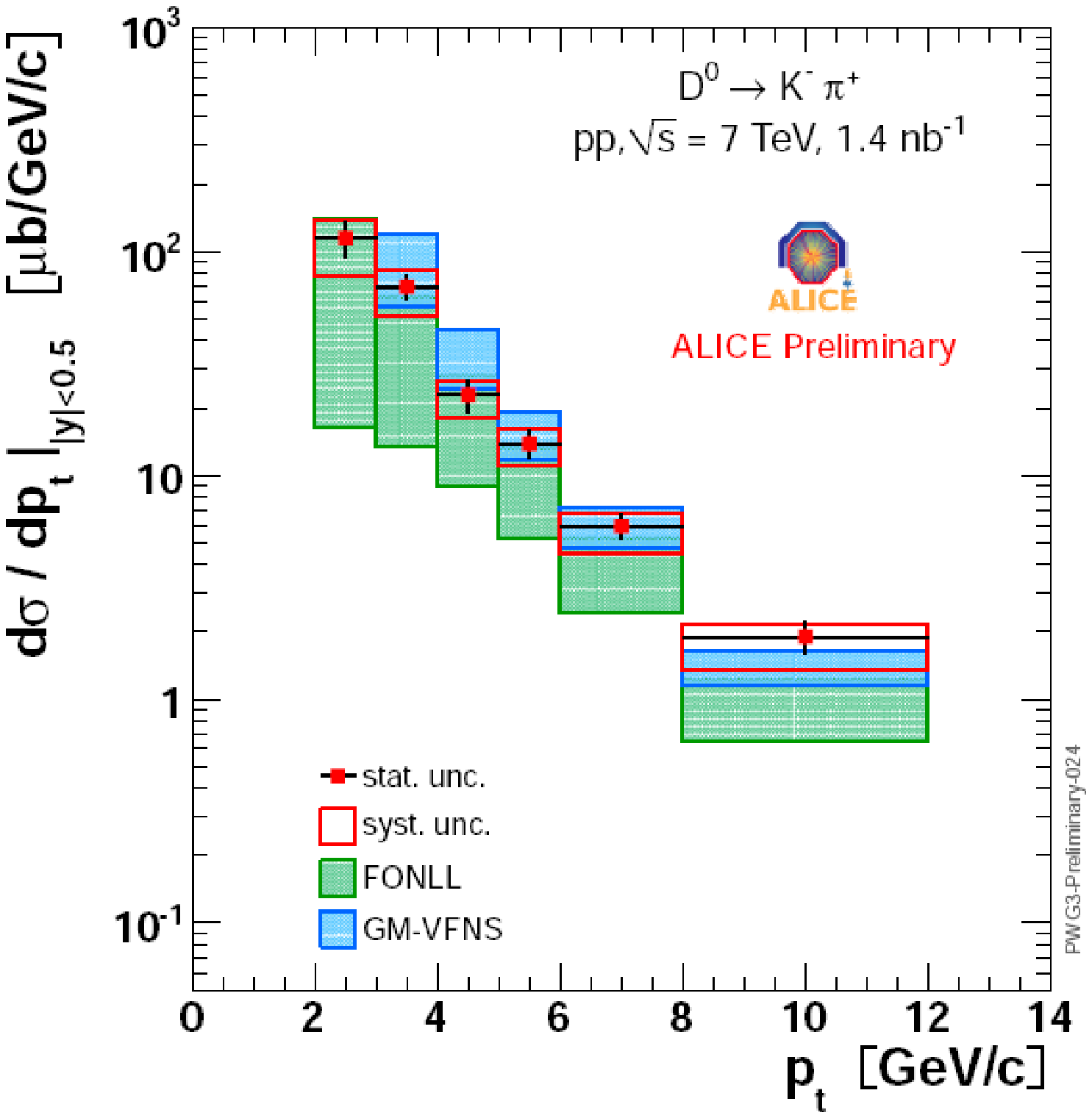}
\hspace{0.2cm}
\includegraphics[height=4.4cm]{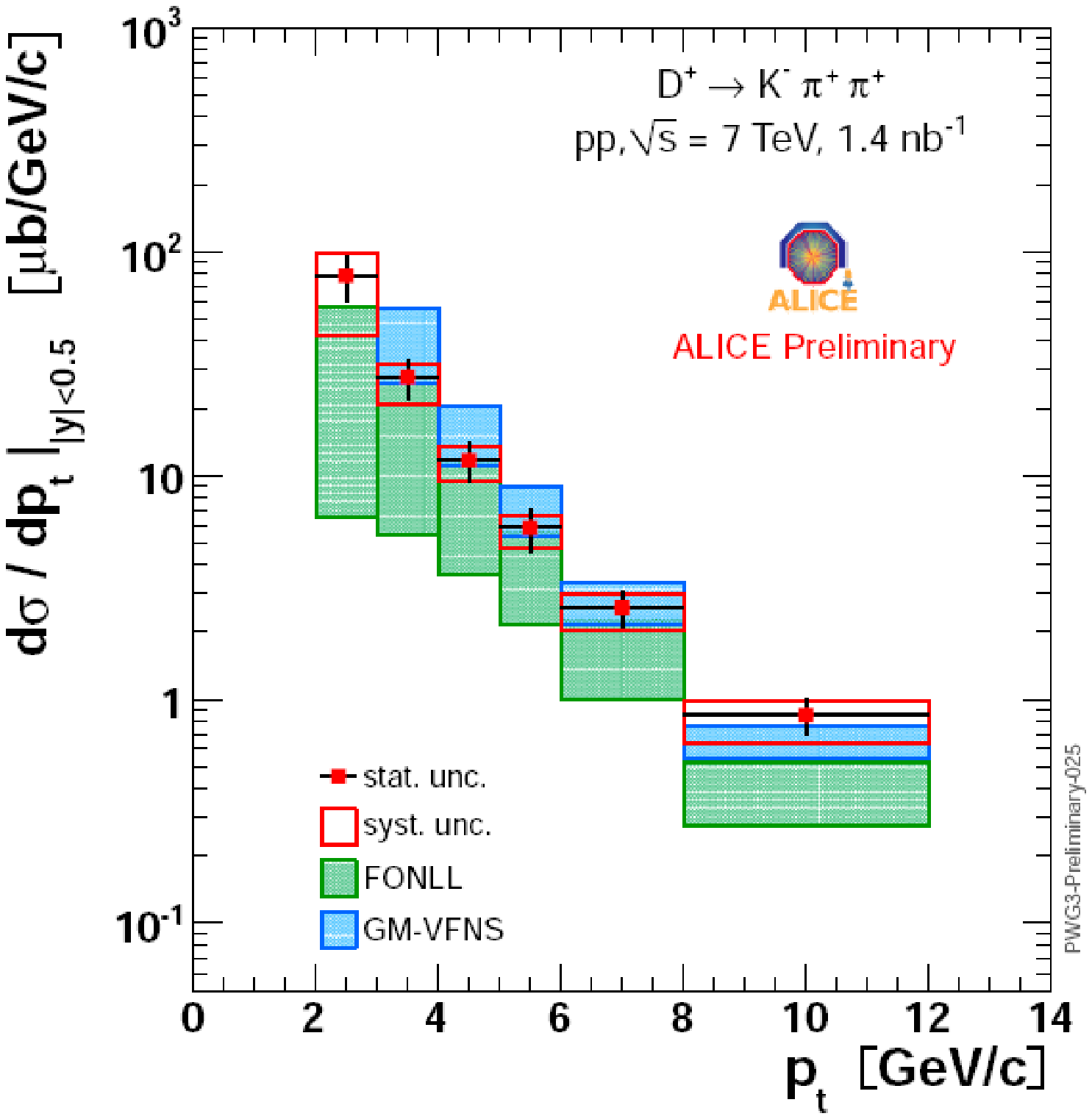}
\hspace{0.2cm}
\includegraphics[height=4.4cm]{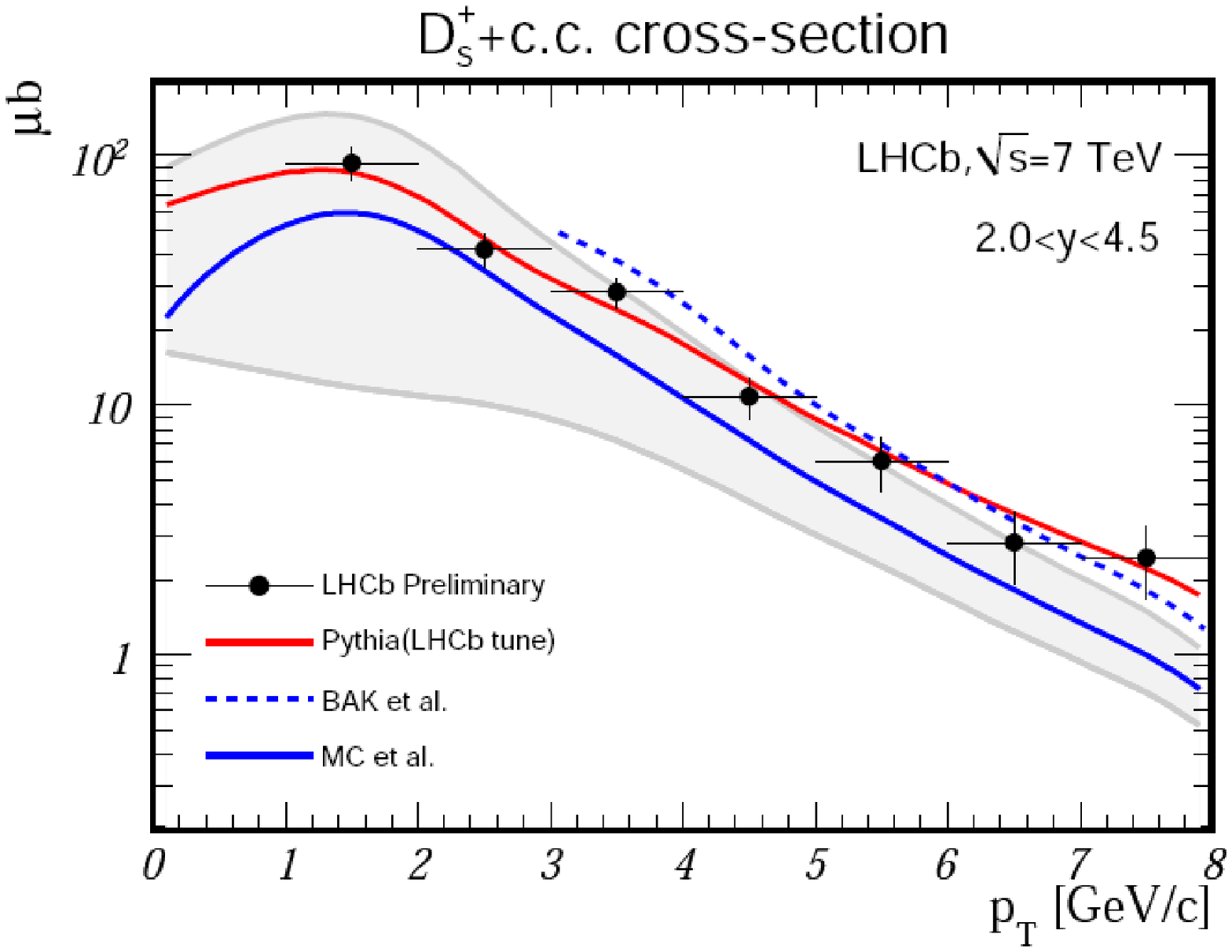}
\caption{\label{fig:LHCOpenFlavour}
$D^0$, $D^+$ and $D_s^+$ hadroproduction at the LHC. GM-VFNS (BAK et al.) and FONLL (MC et al.) predictions are compared to preliminary data measured by the ALICE \cite{ALICEHeavyFlavour} and LHCb \cite{LHCbHeavyFlavour} collaborations. From \cite{ALICEHeavyFlavour} (left and middle) and \cite{LHCbHeavyFlavour} (right).
}
\end{figure}

\begin{figure}
\centering
\includegraphics[height=6cm]{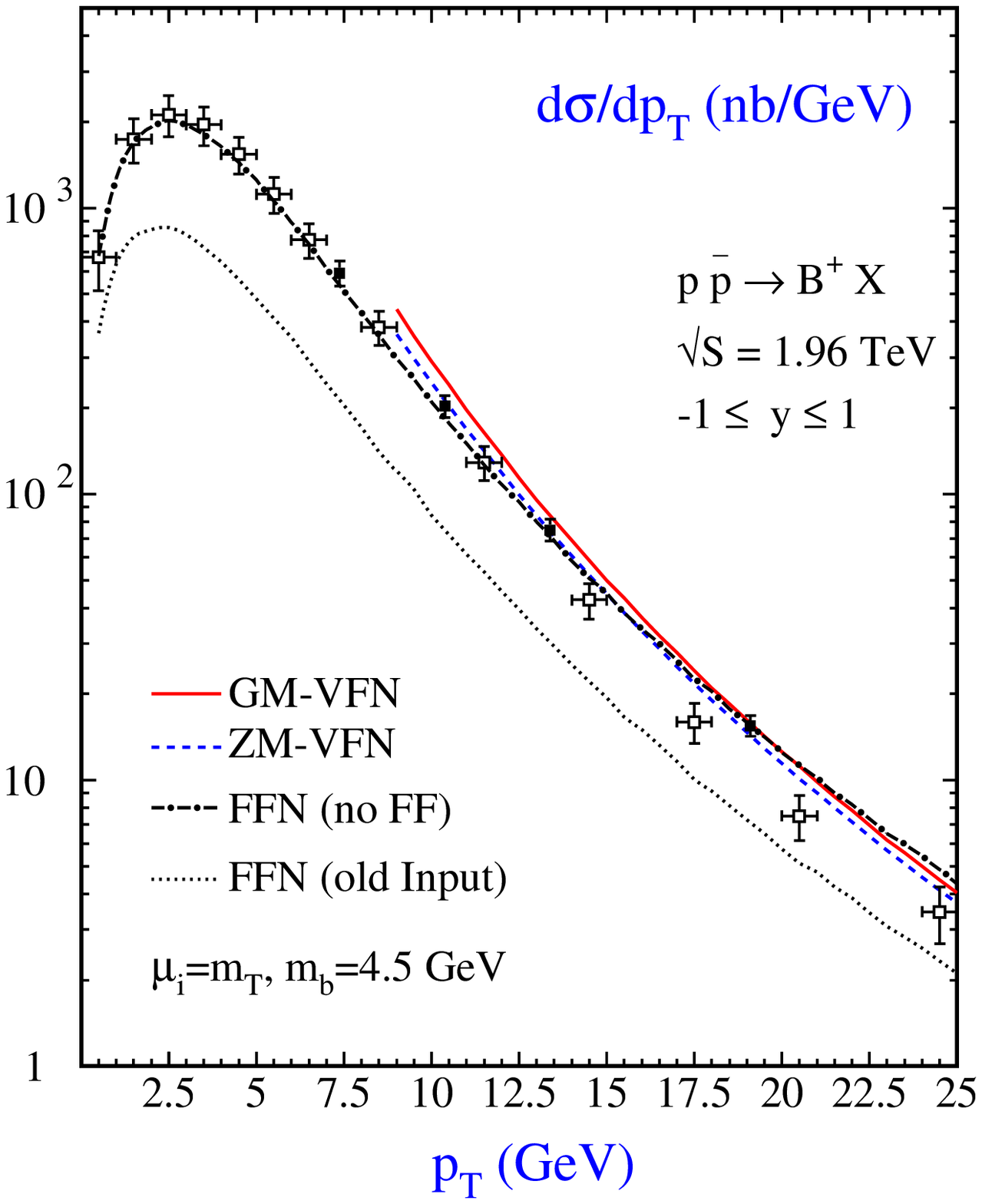}
\hspace{1cm}
\includegraphics[height=6cm]{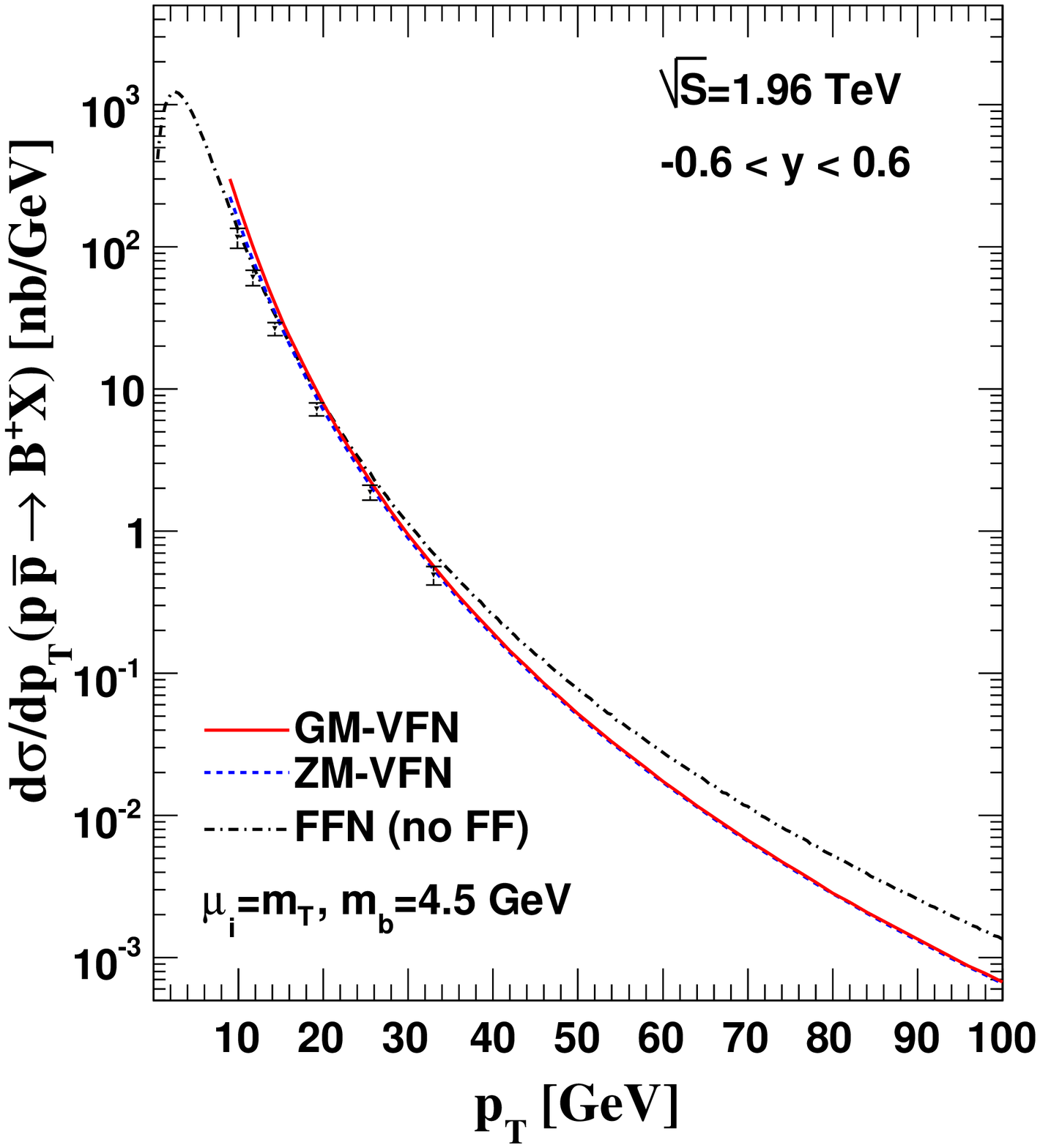}
\caption{\label{fig:BOpenFlavour}
Hadroproduction of $B^+$ mesons at the Tevatron. FFNS, ZM-VFNS and GM-VFNS predictions are compared to data measured by the CDF collaboration \cite{Aaltonen:2009xn}. The ''FFN (old input)`` line is an older prediction of the FFNS evaluated with outdated input parameters.  From \cite{Kniehl:2008zza}.
}
\end{figure}

Predictions at NLO accuracy have been made within the GM-VFNS for $D^{\ast\pm}$ production in two photon-collisions \cite{Kramer:2001gd} and photoproduction \cite{Kramer:2003jw}, for hadroproduction of $D^0$, $D^{\ast\pm}$, $D^{\pm}$, $D_s^\pm$ and $\Lambda_c^\pm$ \cite{Kniehl:2004fy,Kniehl:2005ej} and $B$ meson hadroproduction \cite{Kniehl:2008zza}. NLO predictions in the FONLL scheme have been calculated for $D$ meson photoproduction \cite{Cacciari:2001td}, $D^0$, $D^{\ast\pm}$, $D^{\pm}$, $D_s^\pm$ and $\Lambda_c^\pm$ production at the Tevatron \cite{Cacciari:1998it}, $B$ meson production at the Tevatron \cite{Cacciari:2002pa} as well as $D$ and $B$ meson production at RHIC \cite{Cacciari:2005rk}. An important input for all these calculations are the nonperturbative FFs, which are extracted from fits to scaled energy or momentum distributions of heavy flavour production in electron-positron collisions. In the GM-VFNS framework, these have been extracted in \cite{Binnewies:1997xq} and in the FONLL framework in \cite{Cacciari:2002pa,Cacciari:2003zu}.

In figure \ref{fig:LHCOpenFlavour} GM-VFNS and FONLL predictions for hadroproduction of $D^0$, $D^+$ and $D_s^+$ mesons are compared to recent preliminary data by ALICE \cite{ALICEHeavyFlavour} and LHCb \cite{LHCbHeavyFlavour}. The predictions of both models agree with the data.

In figure \ref{fig:BOpenFlavour} predictions for $B^+$ hadroproduction at the Tevatron \cite{Aaltonen:2009xn} are compared with predictions of the FFNS, ZM-VFNS and GM-VFNS \cite{Kniehl:2008zza}. We see that all three predictions are compatible with the data in their respective regions of applicability. At very high $p_T$ we see the difference between the FFNS on the one hand and the ZM-VFNS and the GM-VFNS on the other hand. At high $p_T$ the latter two agree by construction, while we start to see differences between them at moderate values of $p_T$.

\acknowledgements{%
The author would like to thank Bernd Kniehl for collaboration on our common work about $J/\psi$ production and Bernd Kniehl and Gustav Kramer for advice on the part about open heavy flavour production, which is based on the recent review \cite{Kramer:2011zzb}.
}


%


\begin{thebibliography}{99}
  
\bibitem{Bodwin:1994jh}
  G.~T.~Bodwin, E.~Braaten and G.~P.~Lepage,
  Phys.\ Rev.\  D {\bf 51}, 1125 (1995)
  [Erratum-ibid.\  D {\bf 55}, 5853 (1997)].

\bibitem{Petrelli:1997ge}
  A.~Petrelli, M.~Cacciari, M.~Greco, F.~Maltoni, and M.~L.~Mangano,
  Nucl.\ Phys.\  B {\bf 514}, 245 (1998);
  F.~Maltoni, M.~L.~Mangano and A.~Petrelli,
  Nucl.\ Phys.\  B {\bf 519}, 361 (1998).
  
\bibitem{Klasen:2004az}
  M.~Klasen, B.~A.~Kniehl, L.~N.~Mihaila and M.~Steinhauser,
  Phys.\ Rev.\  D {\bf 71}, 014016 (2005).

\bibitem{Butenschoen:2009zy}
  M.~Butenschoen and B.~A.~Kniehl,
  Phys.\ Rev.\ Lett.\  {\bf 104}, 072001 (2010).

\bibitem{Zhang:2009ym}
  Y.-J.~Zhang, Y.-Q.~Ma, K.~Wang, and K.-T.~Chao,
  Phys.\ Rev.\ D {\bf 81}, 034015 (2010).

\bibitem{Gong:2008ft}
  B.~Gong, X.~Q.~Li and J.~X.~Wang,
  Phys.\ Lett.\  B {\bf 673}, 197 (2009)
  [Erratum-ibid.\  {\bf 693}, 612 (2010)];
  B.~Gong, J.~X.~Wang and H.~F.~Zhang,
  Phys.\ Rev.\  D {\bf 83}, 114021 (2011).

\bibitem{Ma:2010yw}
  Y.-Q.~Ma, K.~Wang, and K.-T.~Chao,
  Phys.\ Rev.\ Lett.\  {\bf 106}, 042002 (2011).

\bibitem{Butenschoen:2010rq}
  M.~Butenschoen and B.~A.~Kniehl,
  Phys.\ Rev.\ Lett.\  {\bf 106}, 022003 (2011).

\bibitem{Butenschoen:2011yh}
  M.~Butenschoen and B.~A.~Kniehl,
  Phys.\ Rev.\ D {\bf 84}, 051501(R) (2011)
  [arXiv:1105.0820 [hep-ph]].


\bibitem{Acosta:2004yw}
  D.~Acosta {\it et al.}\ [CDF Collaboration],
  Phys.\ Rev.\  D {\bf 71}, 032001 (2005).

\bibitem{ALICEdata}
  E.~Scomparin for the ALICE collaboration,
  Nucl.\ Phys.\ B (Proc.\ Suppl.) {\bf 214}, 56 (2011).

\bibitem{ATLASdata}
  G.~Aad {\it et al.}\ [ATLAS Collaboration], ATLAS-CONF-2010-062 (2010).

\bibitem{:2009nj}
  P.~Pakhlov {\it et al.}\ [Belle Collaboration],
  Phys.\ Rev.\  D {\bf 79}, 071101 (2009).

\bibitem{Abe:1997jz}
  F.~Abe {\it et al.}\ [CDF Collaboration],
  Phys.\ Rev.\ Lett.\  {\bf 79}, 572 (1997);
  F.~Abe {\it et al.}  [CDF Collaboration],
  Phys.\ Rev.\ Lett.\ {\bf 79},  578 (1997).

\bibitem{Khachatryan:2010yr}
  V.~Khachatryan {\it et al.}\ [CMS Collaboration],
  Eur.\ Phys.\ J.\  C {\bf 71}, 1575 (2011).

\bibitem{Abdallah:2003du}
  J.~Abdallah {\it et al.}\ [DELPHI Collaboration],
  Phys.\ Lett.\  B {\bf 565}, 76 (2003).

\bibitem{Aaij:2011jh}
  R.~Aaij {\it et al.}\  [LHCb Collaboration],
  Eur.\ Phys.\ J.\  C {\bf 71}, 1645 (2011).

\bibitem{Adare:2009js}
  A.~Adare {\it et al.}\ [PHENIX Collaboration], 
  Phys.\ Rev.\  D {\bf 82}, 012001 (2010).

\bibitem{Chekanov:2002at}
  S.~Chekanov {\it et al.}\ [ZEUS Collaboration],
  Eur.\ Phys.\ J.\  C {\bf 27}, 173 (2003).

\bibitem{Adloff:2002ex}
  C.~Adloff {\it et al.}\ [H1 Collaboration],
  Eur.\ Phys.\ J.\  C {\bf 25}, 25 (2002).

\bibitem{Aaron:2010gz}
  F.~D.~Aaron {\it et al.}\ [H1 Collaboration],
  Eur.\ Phys.\ J.\ C {\bf 68}, 401 (2010).

\bibitem{Artoisenet:2009xh}
  P.~Artoisenet, J.~Campbell, F.~Maltoni, and F.~Tramontano,
  Phys.\ Rev.\ Lett.\  {\bf 102}, 142001 (2009);
  C.~H.~Chang, R.~Li, and J.~X.~Wang,
  Phys.\ Rev.\  D {\bf 80}, 034020 (2009).

\bibitem{Gong:2008sn}
  B.~Gong and J.~X.~Wang,
  Phys.\ Rev.\ Lett.\  {\bf 100}, 232001 (2008);
  B.~Gong and J.~X.~Wang,
  Phys.\ Rev.\  D {\bf 78}, 074011 (2008).

\bibitem{PhotoPolLetter}
  M.~Butenschoen and B.~A.~ Kniehl, arXiv:1109.1476 [hep-ph].

\bibitem{:2009br}
  S.~Chekanov {\it et al.}  [ZEUS Collaboration],
  JHEP {\bf 0912}, 007 (2009).

\bibitem{Beneke:1998re}
  M.~Beneke, M.~Kr\"amer and M.~V\"anttinen,
  Phys.\ Rev.\  D {\bf 57}, 4258 (1998).

\bibitem{Abulencia:2007us}
  A.~Abulencia {\it et al.}  [CDF Collaboration],
  Phys.\ Rev.\ Lett.\  {\bf 99}, 132001 (2007).

\bibitem{Baranov:2002cf}
  S.~P.~Baranov,
  Phys.\ Rev.\  D {\bf 66}, 114003 (2002).

\bibitem{Cho:1994ih}
  P.~L.~Cho and M.~B.~Wise,
  Phys.\ Lett.\  B {\bf 346}, 129 (1995);
  M.~Beneke and M.~Kr\"amer,
  Phys.\ Rev.\  D {\bf 55}, 5269 (1997);
  E.~Braaten, B.~A.~Kniehl and J.~Lee,
  Phys.\ Rev.\  D {\bf 62}, 094005 (2000).
  
\bibitem{Gribov:1984tu}
  L.~V.~Gribov, E.~M.~Levin and M.~G.~Ryskin,
  Phys.\ Rept.\  {\bf 100}, 1 (1983);
%
  E.~M.~Levin and M.~G.~Ryskin,
  Phys.\ Rept.\  {\bf 189}, 267 (1990);
  %
  S.~Catani, M.~Ciafaloni and F.~Hautmann,
  Phys.\ Lett.\  B {\bf 242}, 97 (1990);
%
  S.~Catani, M.~Ciafaloni and F.~Hautmann,
  Nucl.\ Phys.\  B {\bf 366}, 135 (1991);
%
  J.~C.~Collins and R.~K.~Ellis,
  Nucl.\ Phys.\  B {\bf 360}, 3 (1991).

\bibitem{Gribov:1972ri}
  V.~N.~Gribov and L.~N.~Lipatov,
  Sov.\ J.\ Nucl.\ Phys.\  {\bf 15}, 438 (1972);
%
  G.~Altarelli and G.~Parisi,
  Nucl.\ Phys.\  B {\bf 126}, 298 (1977);
%
  Y.~L.~Dokshitzer,
  Sov.\ Phys.\ JETP {\bf 46}, 641 (1977).

\bibitem{Kuraev:1977fs}
  E.~A.~Kuraev, L.~N.~Lipatov and V.~S.~Fadin,
  Sov.\ Phys.\ JETP {\bf 45}, 199 (1977);
%
  I.~I.~Balitsky and L.~N.~Lipatov,
  Sov.\ J.\ Nucl.\ Phys.\  {\bf 28}, 822 (1978)
  [Yad.\ Fiz.\  {\bf 28}, 1597 (1978)].

\bibitem{Ciafaloni:1987ur}
  M.~Ciafaloni,
  Nucl.\ Phys.\  B {\bf 296}, 49 (1988);
  S.~Catani, F.~Fiorani and G.~Marchesini,
  Phys.\ Lett.\  B {\bf 234}, 339 (1990);
  S.~Catani, F.~Fiorani and G.~Marchesini,
  Nucl.\ Phys.\  B {\bf 336}, 18 (1990);
  G.~Marchesini,
  Nucl.\ Phys.\  B {\bf 445}, 49 (1995).
  
\bibitem{Lipatov:2002tc}
  A.~V.~Lipatov and N.~P.~Zotov,
  Eur.\ Phys.\ J.\  C {\bf 27}, 87 (2003);
  S.~P.~Baranov and N.~P.~Zotov,
  J.\ Phys.\ G {\bf 29}, 1395 (2003);
  S.~P.~Baranov, A.~V.~Lipatov and N.~P.~Zotov,
  Eur.\ Phys.\ J.\  C {\bf 71}, 1631 (2011).

\bibitem{Kniehl:2006sk}
  B.~A.~Kniehl, D.~V.~Vasin and V.~A.~Saleev,
  Phys.\ Rev.\  D {\bf 73}, 074022 (2006).

\bibitem{Kniehl:2006vm}
  B.~A.~Kniehl, V.~A.~Saleev and D.~V.~Vasin,
  Phys.\ Rev.\  D {\bf 74}, 014024 (2006).

 \bibitem{Baranov:2007dw}
  S.~P.~Baranov and A.~Szczurek,
  Phys.\ Rev.\  D {\bf 77}, 054016 (2008).

\bibitem{Jung:2000hk}
  H.~Jung and G.~P.~Salam,
  Eur.\ Phys.\ J.\  C {\bf 19}, 351 (2001);
  H.~Jung,
  Comput.\ Phys.\ Commun.\  {\bf 143}, 100 (2002).

\bibitem{Artoisenet:2008fc}
  P.~Artoisenet, J.~M.~Campbell, J.~P.~Lansberg, F.~Maltoni and F.~Tramontano,
  Phys.\ Rev.\ Lett.\  {\bf 101}, 152001 (2008);
  J.~P.~Lansberg,
  Eur.\ Phys.\ J.\  C {\bf 61}, 693 (2009).

\bibitem{Kniehl:2004fy}
  B.~A.~Kniehl, G.~Kramer, I.~Schienbein and H.~Spiesberger,
  Phys.\ Rev.\  D {\bf 71}, 014018 (2005);
  B.~A.~Kniehl, G.~Kramer, I.~Schienbein and H.~Spiesberger,
  Eur.\ Phys.\ J.\  C {\bf 41}, 199 (2005).


\bibitem{Cacciari:1998it}
  M.~Cacciari, M.~Greco and P.~Nason,
  JHEP {\bf 9805}, 007 (1998).
      
\bibitem{Collins:1998rz}
  J.~C.~Collins,
  Phys.\ Rev.\  D {\bf 58}, 094002 (1998).
  
\bibitem{Kramer:2001gd}
  G.~Kramer and H.~Spiesberger,
  Eur.\ Phys.\ J.\  C {\bf 22}, 289 (2001);
  G.~Kramer and H.~Spiesberger,
  Eur.\ Phys.\ J.\  C {\bf 28}, 495 (2003).

\bibitem{Kramer:2003jw}
  G.~Kramer and H.~Spiesberger,
  Eur.\ Phys.\ J.\  C {\bf 38}, 309 (2004);
  B.~A.~Kniehl, G.~Kramer, I.~Schienbein and H.~Spiesberger,
  Eur.\ Phys.\ J.\  C {\bf 62}, 365 (2009).

\bibitem{Kniehl:2005ej}
  B.~A.~Kniehl, G.~Kramer, I.~Schienbein and H.~Spiesberger,
  Phys.\ Rev.\ Lett.\  {\bf 96}, 012001 (2006);
  B.~A.~Kniehl, G.~Kramer, I.~Schienbein and H.~Spiesberger,
  Phys.\ Rev.\  D {\bf 79}, 094009 (2009).

\bibitem{Kniehl:2008zza}
  B.~A.~Kniehl, G.~Kramer, I.~Schienbein and H.~Spiesberger,
  Phys.\ Rev.\  D {\bf 77}, 014011 (2008).
  
\bibitem{Cacciari:2001td}
  M.~Cacciari, S.~Frixione and P.~Nason,
  JHEP {\bf 0103}, 006 (2001).
  
\bibitem{Cacciari:2002pa}
  M.~Cacciari and P.~Nason,
  Phys.\ Rev.\ Lett.\  {\bf 89}, 122003 (2002);
  M.~Cacciari, S.~Frixione, M.~L.~Mangano, P.~Nason and G.~Ridolfi,
  JHEP {\bf 0407}, 033 (2004).
  
\bibitem{Cacciari:2005rk}
  M.~Cacciari, P.~Nason and R.~Vogt,
  Phys.\ Rev.\ Lett.\  {\bf 95}, 122001 (2005).
  
\bibitem{Binnewies:1997xq}
  J.~Binnewies, B.~A.~Kniehl and G.~Kramer,
  Phys.\ Rev.\  D {\bf 58}, 014014 (1998);
  J.~Binnewies, B.~A.~Kniehl and G.~Kramer,
  Phys.\ Rev.\  D {\bf 58}, 034016 (1998);
  B.~A.~Kniehl and G.~Kramer,
  Phys.\ Rev.\  D {\bf 71}, 094013 (2005);
  B.~A.~Kniehl and G.~Kramer,
  Phys.\ Rev.\  D {\bf 74}, 037502 (2006);
  T.~Kneesch, B.~A.~Kniehl, G.~Kramer and I.~Schienbein,
  Nucl.\ Phys.\  B {\bf 799}, 34 (2008).
    
\bibitem{Cacciari:2003zu}
  M.~Cacciari and P.~Nason,
  JHEP {\bf 0309}, 006 (2003);
  M.~Cacciari, P.~Nason and C.~Oleari,
  JHEP {\bf 0604}, 006 (2006).
  
\bibitem{ALICEHeavyFlavour}
  A.~Dainese, presented at the LHC physics day ''Charm and bottom quark production at the LHC``, 3 December 2010, CERN.

\bibitem{LHCbHeavyFlavour}
  P.~Urquijo for the LHCb collaboration, LHCb-CONF-2010-013 (2010).

\bibitem{Aaltonen:2009xn}
  T.~Aaltonen {\it et al.}  [CDF Collaboration],
  Phys.\ Rev.\  D {\bf 79}, 092003 (2009).

\bibitem{Kramer:2011zzb}
  G.~Kramer,
  Nucl.\ Phys.\ Proc.\ Suppl.\  {\bf 214}, 123 (2011).
  
\end{thebibliography}
\end{document}